\documentclass[
aip,
amsmath,amssymb,
preprint,%
 reprint,%
]{revtex4-1}

\usepackage{graphicx}
\usepackage{xcolor}
\usepackage{CJKutf8}
\usepackage{dcolumn}
\usepackage{bm}

\usepackage[utf8]{inputenc}
\usepackage[T1]{fontenc}
\usepackage{mathptmx}
\usepackage{etoolbox}
\usepackage{adjustbox}
\usepackage{here}

\makeatletter
\def\@email#1#2{%
 \endgroup
 \patchcmd{\titleblock@produce}
  {\frontmatter@RRAPformat}
  {\frontmatter@RRAPformat{\produce@RRAP{*#1\href{mailto:#2}{#2}}}\frontmatter@RRAPformat}
  {}{}
}%
\makeatother
\begin{document}

\preprint{AIP/123-QED}

\title{
Analytical and Numerical Linear Analyses of Convection Revisited
}
\author{Izumi Seno
}
\email{seno.izumi.y0@s.mail.nagoya-u.ac.jp}
\affiliation{ 
Department of Physics, Graduate School of Science, Nagoya University, Furo-cho, Chikusa-ku, Nagoya, Aichi 464-8602, Japan
}
\author{Shu-ichiro Inutsuka
}
\affiliation{ 
Department of Physics, Graduate School of Science, Nagoya University, Furo-cho, Chikusa-ku, Nagoya, Aichi 464-8602, Japan
}
\date{\today}

\begin{abstract}
We conduct linear analyses of convection in domains larger than the temperature scale height. 
We employ both analytical and numerical methods in these analyses. 
In the case excluding all dissipation, the typical time scale of convection is determined by the free fall time over the temperature scale height.
We quantitatively show the condition for the Boussinesq and Wentzel-Kramers-Brillouin (WKB) approximations to be applicable.
We provide a reassessment of the critical Rayleigh number, a key indicator of convection, and show that WKB approximation tends to underestimate the critical Rayleigh number, particularly when the temperature scale height is comparable to or smaller than the domain height.
We show clear explanation why both thermal conduction and viscosity are required for stabilizing negative entropy gradient medium.
\end{abstract}

\maketitle

\section{Introduction}
\label{sec:introduction}

Convection is driven by buoyancy forces in a gravitationally stratified medium and plays a crucial role in numerous astrophysical and geophysical processes, such as stellar convection zones \cite{Chandrasekhar1981}, atmospheric dynamics \cite{Holton1973}, and oceanic circulation \cite{Pedlosky1987}. 
The theoretical study of convection has a long history\cite{annurev}.
For example, Chandrasekhar has provided comprehensive analyses of various situations, including the effects of rotation\cite{Takehiro_etal2002} and magnetic fields \cite{Chandrasekhar1981}. 
While many important contributions have been made, dispersion relations of convective instabilities in general conditions are not available, particularly in systems with strong temperature contrasts in a gravitationally stratified environment.  

The Wentzel-Kramers-Brillouin (WKB) approximation has been widely used to derive dispersion relations in convective systems, particularly when the temperature scale height is significantly larger than the domain height \cite{Gough1969, Kaladze_Misra2024}. 
However, the WKB approximation assumes sinusoidal or exponential functions as the perturbation profiles, but those functions cannot be good approximations of the eigenfunctions for the unperturbed states with large gradients.
There is no convenient analytic profile for eigenfunctions for inhomogeneous system, which makes WKB approximation impracticable.

The Boussinesq approximation is another commonly used approximation that simplifies the governing equations by neglecting sound wave modes. 
Under this approximation, the buoyancy force can be estimated only by the temperature perturbation, which leads to the definition of the Rayleigh number, a non-dimensional quantity that is often used to characterize the stability of the system:
\begin{align}
R{\rm a} = \dfrac{g \beta L_z^4}{T \kappa \nu},
\label{eq:rayleigh_number}
\end{align}
where $g$, $\kappa$, and $\nu$ represent gravity, thermal diffusivity, and kinematic viscosity, respectively \cite{Spiegel1962}. 
In addition, $T$ represents the temperature, $\beta \equiv -dT/dz$ denotes the temperature gradient, and $L_z$ is the domain height. 
The reason for the positivity of $\beta$ will be discussed in Section \ref{sec:ScaleHeight}.
When the temperature gradient is spatially constant, as considered in this paper, Equation \eqref{eq:rayleigh_number} can be rewritten as $R{\rm a} = g\Delta T L_z^3 / T\kappa \nu$, where $\Delta T$ is the temperature difference between the top and bottom boundaries \cite{Chandrasekhar1981, Turcotte2014}.
Here, $T$ represents the characteristic temperature of the system, such as the temperature of the mid-point in the case of a linear temperature profile, as in Equation \eqref{eq:Unp_temp_distribution}.
Although the Rayleigh number provides a measure of the balance between buoyancy forces and the effects of diffusion, its interpretation may become more complicated when the temperature scale height is comparable to or smaller than the domain height. 
Furthermore, the limits of the Boussinesq approximation in such cases continue to be the subject of ongoing research \cite{Landau1987}.

Recent studies have explored the relationship between convection and internal gravity waves, as well as classical problems such as Rayleigh-B\`{e}nard convection\cite{Tiantian_Jung_2025, Dowling_1988} driven by temperature differences between horizontal layers \cite{Turcotte2014}. 
While these studies offer valuable insights, they often do not provide a precise dispersion relation that accounts for the influence of the background structure. 
For example, Kaladze and Misra \cite{Kaladze_Misra2024} have contributed valuable perspectives on dispersion relations, but have not addressed the effects of strong inhomogeneities on the validity of the WKB approximation.

In this study, we perform linear analyses of convective instability using both analytical and numerical methods. 
Our findings suggest that the Boussinesq approximation remains valid when the sound crossing time over the $temperature$ scale height exceeds the free-fall time over the $temperature$ scale height. 
However, the WKB approximation tends to overestimate both the growth rate and the critical Rayleigh number when the $temperature$ scale height is smaller than the system height. 
Note that the Boussinesq and WKB approximations are totally different concepts and independent of each other.
By performing numerical linear analyses, we refine these estimations and demonstrate that the WKB approximation holds only when the $temperature$ scale height is significantly larger than the domain height.

The structure of this paper is as follows. 
In Section \ref{sec:ScaleHeight}, we discuss various scale heights that are important for convective systems.
Section \ref{sec:method} outlines our linear perturbation analyses as the eigenvalue problem by setting up the unperturbed background state and derive the linearized perturbation equations.
Section \ref{sec:analytical} presents the analytical approach, utilizing classical approximations. 
In Section \ref{sec:numerical}, we conduct a numerical linear analysis using the shooting method. 
Section \ref{sec:results} compares analytical and numerical results, and Section \ref{sec:discussion} provides a detailed discussion of the effects of diffusion, the requirement for stability of the medium with negative entropy gradient, the critical Rayleigh number as a function of temperature scale height, and the applicability of the Boussinesq approximation. 
Finally, we summarize our findings in Section \ref{sec:summary}.

\section{Key Scale Heights in Convective Systems}
\label{sec:ScaleHeight}
In this section, we distinguish various scale heights in the gravitationally stratified medium.
We use Cartesian coordinates $(x, y, z)$ and a symbol $g (>0)$ as the constant gravitational acceleration in the vertical downward direction (i.e., $-z$ direction).
In inhomogeneous systems, scale heights can be defined along the $z-$direction.
\begin{align}
    \dfrac{1}{H_P} &\equiv -\dfrac{1}{P}\dfrac{dP}{dz} = \dfrac{\rho g}{P} > 0,
    \label{eq:intro:HP}\\[3mm]
    \dfrac{1}{H_\rho} &\equiv -\dfrac{1}{\rho}\dfrac{d\rho}{dz} = \dfrac{1}{H_P} + \dfrac{1}{T}\dfrac{dT}{dz} > 0,
    \label{eq:intro:Hrho}\\[3mm]
    \dfrac{1}{H_s} &\equiv -\dfrac{1}{P/\rho^\gamma}\dfrac{d(P/\rho^\gamma)}{dz} = (1 - \gamma)\dfrac{1}{H_P} - \gamma\dfrac{1}{T}\dfrac{dT}{dz} > 0,
    \label{eq:intro:Hs}
\end{align}
where $H_P$, $H_\rho$, and $H_s$ represent the scale heights of pressure ($P$), density ($\rho$), and entropy ($\log s \equiv P/\rho^\gamma$, where $s$ denotes entropy), respectively. 
Here, $\gamma$ denotes the heat capacity ratio, and $T$ is the temperature. 
The positivity of Equation \eqref{eq:intro:Hrho} is required by Rayleigh-Taylor stability\citep{Chandrasekhar1981}.
It is important to note that the fundamental relationship facilitating the derivation of these equations is given by the following equation:
\begin{align*}
    \dfrac{dP}{P} = \dfrac{d\rho}{\rho} + \dfrac{dT}{T}.
\end{align*}
The positivity of Equation \eqref{eq:intro:Hs} means negative entropy gradient\footnote{This is equivalent to the condition that the square of the $\mathrm{Brunt-V\ddot{a}is\ddot{a}i\ddot{a}}$ frequency is negative, i.e., $\omega_{\rm BV}^2 = (g/\rho C_{\rm s}^2) \times [(dP_0/dz) - C_{\rm s}^2 (d\rho_0/dz)] < 0$.} and necessary condition for convection \cite{Landau1987, Mihalas1999}.
In general, the specific heat ratio $\gamma$ satisfies $\gamma \geq 1$. 
Consequently, the temperature gradient, $dT / dz$, must be negative as a necessary condition for convective instability. 
Since we define $\beta \equiv -dT / dz$, this condition can be rephrased as $\beta > 0$. 
According to this definition, the temperature scale height, $H_T$, is defined as:
\begin{align}
    \dfrac{1}{H_T} \equiv \dfrac{\beta}{T} = -\dfrac{1}{T}\dfrac{dT}{dz} 
    > (\gamma - 1)\dfrac{1}{H_\rho} > 0,
    \label{eq:intro:HT}
\end{align}
where $H_\rho > (\gamma - 1)H_T$ is derived from the condition of convective instability, $H_s > 0$.
Finally, the inverse of the pressure scale height, normalized by the height of the computational domain, $L_z$, can be expressed as the sum of the inverse scale heights of density and temperature:
\begin{align}
    \dfrac{L_z}{H_P} = \dfrac{L_z}{H_\rho} + \dfrac{L_z}{H_T}.
    \label{eq:intro:HP_sum}
\end{align}
Since we are considering a hydrostatic equilibrium state, where $H_P > 0$ and $H_\rho > 0$, the pressure scale height must be smaller than the temperature scale height, i.e., $H_T > H_P$, as indicated by Equation \eqref{eq:intro:HP_sum}.
Consequently, if the temperature scale height is smaller than the domain height, the pressure scale height will also be smaller than the domain height.

In the case of large pressure scale height, $H_P > L_z$, both density and temperature scale heights are larger than the domain height because of Equation \eqref{eq:intro:HP_sum} and the positivity of $H_\rho$ and $H_T$
(Note that this relation is not always expected for the system of $\beta < 0,\ H_T < 0$ that is stable for convection, $H_s < 0$.).
This condition corresponds to a state where all of $\rho, T, P$, and $s$, are only slowly decrease along the vertical direction of the computational domain, which may allow WKB approximation for nearly homogeneous system.
Indeed, in later sections, we will see that WKB approximation for the case of $H_P \gg L_z$ provides accurate result.

The main objective of this article is the opposite case.  
According to Equation \eqref{eq:intro:HP_sum}, when either the density or temperature scale height is smaller than the domain height, the pressure scale height must be smaller than the domain height.  
We identify only two possibilities in this case for convection:
\begin{enumerate}
    \item $H_\rho > L_z$ and $H_T < L_z$.
    
    The temperature scale height is smaller than the domain height, while the density scale height is not. 
    In this case, pressure scale height is smaller than the domain height.
    \item $H_\rho < L_z$ and $H_T < L_z$.
    
    Both the density and temperature scale heights are smaller than the domain height, and hence the pressure scale height is smaller than the domain scale height. 
\end{enumerate}
As mentioned earlier, the system remains convectively stable, i.e., $H_s < 0$, when the density scale height is smaller than the domain height while the temperature scale height is not, i.e., $H_\rho < L_z,\ H_T > L_z$.  
In the later sections, we will see that not only Case 2 but also Case 1 show deviation from the results obtained by WKB approximation. 
Hereafter we mostly focus on Case 1.

\section{Formulation of Eigenvalue problem}
\label{sec:method}
We consider the dynamics of an ideal gas under the influence of gravity and viscosity. To describe this system, we employ the standard hydrodynamic equations, which govern the conservation of mass, momentum, and energy:
\begin{align}
& 
\dfrac{\partial \rho}{\partial t} + \nabla \cdot [\rho \bm{v}] = 0,
\label{eq:Norm_EoC} \\[3mm]
& 
\rho \left [ \dfrac{\partial \bm{v}}{\partial t} + (\bm{v} \cdot \nabla) \bm{v} \right ]
= -\nabla P + \rho \nu \nabla^2 \bm{v} - \rho g \bm{e}_z,
\label{eq:Norm_EoM} \\[3mm]
& 
\dfrac{\partial T}{\partial t} + (\bm{v} \cdot \nabla) T = \kappa \nabla^2 T.
\label{eq:Norm_Energy}
\end{align}
Here, $\rho$, $P$, and $T$ denote density, pressure, and temperature, respectively.
The velocity field is represented by $\bm{v}$ and  $\bm{e}_z$ is the unit vector in the vertical direction.
Additionally, $\nu$ and $\kappa$ represent the viscosity coefficient and thermal diffusivity. 
For the sake of simplicity, these parameters and gravitational acceleration, $g$, don't depend on temperature and temperature\cite{article}, and are treated as constants.
This formulation allows us to analyze the basic behavior of fluid motion in the presence of gravity while taking into account the effects of viscosity and thermal diffusion.

\subsection{Unperturbed state}
\label{subsec:Unp}
In this section, we define the unperturbed state in preparation for the linear analysis in the subsequent section.
The unperturbed state variables are indicated by a subscript ``0". 
We assume a horizontally uniform hydrostatic density profile.
The unperturbed state is then given by the following one-dimensional ordinary differential equations (ODEs):
\begin{align}
&
\dfrac{dP_0}{dz} = -\rho_0 g,
\label{eq:Unp_EoM}\\[3mm]
&
\dfrac{d^2 T_0}{dz^2} = 0,
\label{eq:Unp_Energy}\\[3mm]
&
P_0 = \dfrac{R}{\mu}\rho_0 T_0,
\label{eq:Unp_EoS}
\end{align}
where $R$ is the gas constant and $\mu$ is the mean molecular weight.

From Equation (\ref{eq:Unp_Energy}), we can derive the linear temperature distribution as follows:
\begin{align}
T_0(z) = T_{\rm mid} - \beta \left (z - 0.5 L_z \right ),
\label{eq:Unp_temp_distribution}
\end{align}
where $T_{\rm mid}$ is the temperature at the mid-plane ($z/L_z = 0.5$) and $\beta$ is constant throughout the entire domain.
In our analysis, the temperature gradient, $\beta$, is a free parameter, and the temperature scale height is given by the expression, $H_T = T_{\rm mid}/\beta$, which is the definition of the temperature scale height in this paper.

Using the fourth-order Runge-Kutta integrator, we numerically solve Equations (\ref{eq:Unp_EoM}) with (\ref{eq:Unp_EoS}) and (\ref{eq:Unp_temp_distribution}) to obtain the unperturbed profile. 
We solve the ODEs from $z = 0$ to $z = L_z$. 
The bottom boundary condition of pressure is set by specifying 
$
P_0(z=0) / (R T_{\rm mid}/\mu L_z^{3}) = 1
$.

\begin{figure}[t]
\centering
\includegraphics[width=\linewidth]{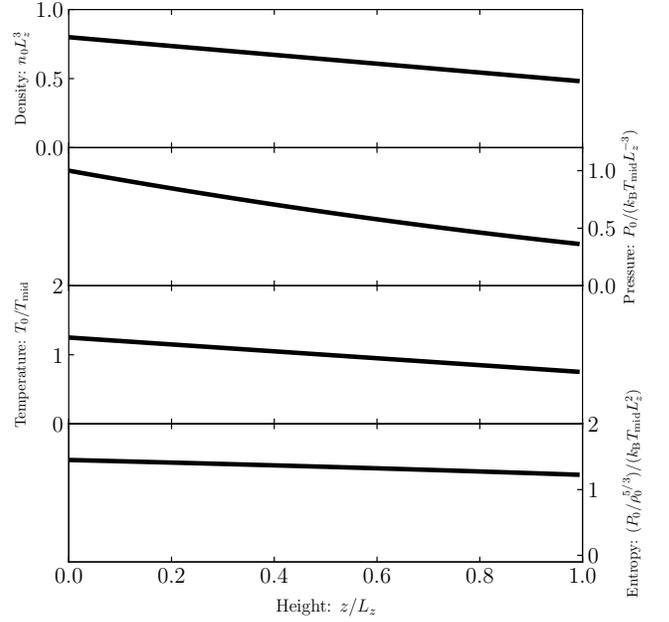}
\caption{
The unperturbed profiles for various physical quantities. 
This figure presents the density, pressure, temperature, and entropy profiles for the case where $H_T / L_z = 2$.
All values are normalized with respect to the mid-plane temperature, gravitational acceleration, and system height.
}
\label{fig:unperturbed}
\end{figure}

Figure \ref{fig:unperturbed} shows the unperturbed profiles obtained by solving Equations (\ref{eq:Unp_EoM})-(\ref{eq:Unp_temp_distribution}) for the case where $H_T / L_z = 2$. 
The bottom panel displays the exponential of the entropy profile, defined as $P_0/\rho_0^{5/3}$,
where the index corresponds to the heat capacity ratio for a monatomic ideal gas. 
This profile has a negative entropy gradient, a characteristic condition for convection\cite{Mihalas1999}.
In the following sections, we will focus on the linear analysis of unperturbed states of $H_T / L_z = 2, 100$.
Note that the unperturbed state does not depend on the values of viscosity coefficient and thermal diffusivity.

\subsection{Linearized equations}
\label{subsec:linear}
In this section, we derive the linearized equations based on Equations (\ref{eq:Norm_EoC}) - (\ref{eq:Norm_Energy}) and (\ref{eq:Unp_EoS}).
We assume that the perturbation of a physical quantity $f$ takes the form $\delta f(t, z) \exp(ik_x x)$, where $k_x$ is the wave number in the horizontal direction. 
Then we can derive the following linearized equations:
\begin{align}
&
\dfrac{\partial \delta \rho}{\partial t} + ik_x \rho_0 \delta v_x + \rho_0 \dfrac{d\delta v_z}{dz} + \delta v_z \dfrac{d\rho_0}{dz} = 0,
\label{eq:linear_EoC}
\\[3mm]
&
\rho_0 \dfrac{\partial \delta v_x}{\partial t} =
-ik_x\delta P - k_x^2 \rho_0 \nu \delta v_x + \rho_0 \nu \dfrac{d^2 \delta v_x}{dz^2},
\label{eq:linear_EoMx}
\\[3mm]
&
 \rho_0 \dfrac{\partial \delta v_z}{\partial t} =
-\dfrac{d\delta P}{dz} - k_x^2 \rho_0 \nu \delta v_z + \rho_0 \nu \dfrac{d^2 \delta v_z}{dz^2} - g\delta \rho,
\label{eq:linear_EoMz}
\\[3mm]
&
\dfrac{\partial \delta T}{\partial t} - \beta \delta v_z = -\kappa k_x^2  \delta T + \kappa \dfrac{d^2 \delta T}{dz^2},
\label{eq:linear_Energy}
\\[3mm]
&
\dfrac{\delta P}{P_0} = \dfrac{\delta \rho}{\rho_0} + \dfrac{\delta T}{T_0}.
\label{eq:linear_EoS}
\end{align}
Since the $x$ and $y$ directions are uniform and governed by the same equations of motion, we will hereafter conduct our analysis in the two-dimensional analysis in the $(x, z)$-plane\cite{Ishiwatari_Takehiro_Hayashi_1994}.
In the three-dimensional analysis $k_x$ is replaced by $k_{\rm h} \equiv \sqrt{k_x^2 + k_y^2}$ and the rest of the analysis remains the same.

\section{Analytical Approach}
\label{sec:analytical}
In this section, we analytically derive the dispersion relations by transforming Equations (\ref{eq:linear_EoC}) - (\ref{eq:linear_EoS}) before performing the numerical analysis.

\subsection{Application of WKB approximation}
\label{subsec:WKB}
In this section, we apply the WKB approximation to analytically solve the ODEs (\ref{eq:linear_EoC})-(\ref{eq:linear_Energy}). 
Assuming the variation in the vertical direction takes the form $\delta f(t, z) = \delta f \exp(\sigma t + ik_z z)$, where $\sigma$ is the growth rate and $k_z$ is the wave number in the vertical direction, we derive the following dispersion relation:
\begin{align}
\begin{aligned}
&
\left( ik_z + \dfrac{\mu g}{R T} \right) \left( ik_z + \dfrac{1}{\rho} \dfrac{d\rho}{dz} \right)
(\sigma + \kappa k^2 ) (\sigma + \nu k^2 )
\\[2mm]
=&
\left\{ k_x^2 + \dfrac{\mu \sigma (\sigma + \nu k^2)}{RT} \right\}
\left\{ (\sigma + \kappa k^2)(\sigma + \nu k^2) - \dfrac{g}{H_T} \right\}
\\[2mm]
&+
\left( ik_z + \dfrac{\mu g}{R T} \right)(\sigma + \nu k^2) \dfrac{\sigma}{H_T},
\end{aligned}
\label{eq:WKB_DR}
\end{align}
where $k^2 = k_x^2 + k_z^2$.

The imaginary part of Equation (\ref{eq:WKB_DR}) is given by the following:
\begin{align}
\dfrac{1}{H_T} (\sigma + \nu k^2) \kappa k^2 k_z = 0.
\label{eq:WKB_DR_Im}
\end{align}
From Equation (\ref{eq:WKB_DR_Im}), we only find the solution for general $k_z$ is $\sigma = - \nu k^2$, but this solution does not satisfy Equation (\ref{eq:WKB_DR}).
This means that there is no solution for a real growth rate with sinusoidal perturbations. This is reasonable because the plane wave solution is not the exact solution for a structure with a gradient \cite{Kaladze_Misra2024}. 
However, as demonstrated in Section \ref{sec:results}, it is possible to obtain a real growth rate for the eigenfunctions whose dependences in the $z$-direction are not sinusoidal. This suggests that the most unstable mode of convection for each $k_x$ is not overstable and has a real growth rate. Thus, we seek for an unstable mode whose growth rate is real even within the WKB approximation.

When the temperature scale height, $1/H_T$, is significantly greater than the domain height, the imaginary part of the dispersion relation (\ref{eq:WKB_DR}) becomes sufficiently small because it has a factor of $1/H_T$. 
Thus, ignoring the imaginary part and considering only the real part of Equation (\ref{eq:WKB_DR}), we obtain the following equation:
\begin{align}
\begin{aligned}
0=\ 
\sigma^4 
&
+ (\kappa + 2 \nu)k^2 \sigma^3 
\\[3mm]
&
+ \left \{ \nu (2\kappa + \nu ) k^4 + \dfrac{RT}{\mu}k^2 + \dfrac{g}{H_\rho}\right \} \sigma^2 
\\[3mm]
&
+ \left \{ \kappa \nu^2 k^6  +\left (\dfrac{RT}{\mu}k^2 + \dfrac{g}{H_\rho} 
\right ) (\kappa + \nu) k^2 \right \} \sigma
\\[3mm]
&
+ \left (\dfrac{RT}{\mu}k^2 + \dfrac{g}{H_\rho}  \right ) \kappa \nu k^4 - \dfrac{Rg\beta }{\mu} k_x^2,
\end{aligned}
\label{eq:WKB_DR_kz0}
\end{align}
This approximation for the solution is valid when the temperature scale height is much larger than the domain height, as will be quantitatively demonstrated in the following section.

Although Equation (\ref{eq:WKB_DR_kz0}) is quite complex, it is noteworthy that all coefficients of $\sigma$, apart from the constant term, are positive. 
Defining the left-hand side of the equation as $f(\sigma)$, the axis of $f(\sigma)$ lies in the region where $\sigma < 0$ on the $\sigma - f(\sigma)$ plane.
This implies that if the constant term (the last line of Equation \eqref{eq:WKB_DR_kz0}) is negative, the mode is unstable. 
The system becomes stable when $k_x = 0$. 
For $k_x \neq 0$, the sufficient condition for instability is described as follows:
\begin{align}
k^6 + \dfrac{\mu g}{RT H_\rho}k^4 - \dfrac{g\beta}{T\kappa \nu}k_x^2 < 0.
\label{eq:WKB_Unstable}
\end{align}

\begin{figure}[ht]
\centering
\includegraphics[width=0.9\linewidth]{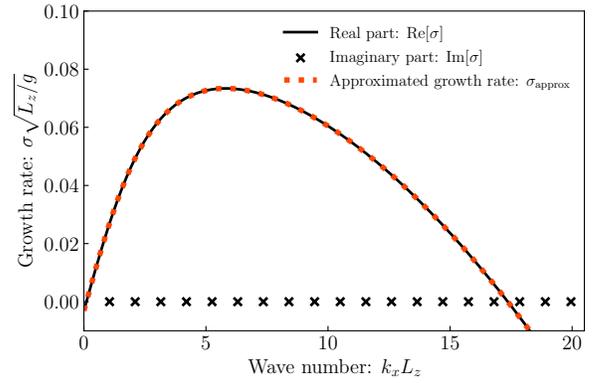}
\caption{
The growth rate as a function of wave number for the case where $H_T / L_z = 100$ and $\kappa / \sqrt{gL_z^3} = \nu / \sqrt{gL_z^3} = 10^{-3.5}$.
The black solid line represents the real part of the growth rate from Equation (\ref{eq:WKB_DR}), while the cross markers indicate the imaginary part. The red dotted line represents the solution of Equation (\ref{eq:WKB_DR_kz0}).
}
\label{fig:WKB_Real_Imagin}
\end{figure}

Then, we should demonstrate the smallness the imaginary part of the growth rate and accuracy of the dispersion relation obtained by solving Equation (\ref{eq:WKB_DR_kz0}). 
Figure \ref{fig:WKB_Real_Imagin} presents the growth rate as a function of the wave number $k_x$, derived from solving Equations (\ref{eq:WKB_DR}) and (\ref{eq:WKB_DR_kz0}) for the case where $H_T / L_z = 100$ and $\kappa / \sqrt{gL_z^3} = \nu / \sqrt{gL_z^3} = 10^{-3.5}$.
It is evident that the real part of the solution from Equation (\ref{eq:WKB_DR}) closely matches the solution from Equation (\ref{eq:WKB_DR_kz0}). 
Although the imaginary part is non-zero, it remains sufficiently small.
Thus, we can discuss the stability of convection by solving only the real part of Equation (\ref{eq:WKB_DR}) expressed as Equation (\ref{eq:WKB_DR_kz0}).

\paragraph{Boussinesq Approximation:} 
Moreover, when we apply the incompressible condition ($\nabla \cdot \bm{v} = 0$) and the Boussinesq approximation ($g \delta \rho \approx -\rho_0 \alpha g \delta T$, where $\alpha = 1/T$), the analytical dispersion relation simplifies as follows:
\begin{align}
k^2 \sigma^2 + (\kappa + \nu) k^4 \sigma + \kappa \nu k^6 - \dfrac{g}{H_T} k_x^2 = 0.
\label{eq:WKB_DR_boussinessq}
\end{align}
In contrast to Equation (\ref{eq:WKB_DR}), this dispersion relation does not include complex terms. Equation (\ref{eq:WKB_DR_boussinessq}) also has an unstable solution when the following criterion is satisfied:
\begin{align}
k^6 - \dfrac{g \beta}{T \kappa \nu} k_x^2 < 0.
\label{eq:WKB_Unstable_boussinessq}
\end{align}
This is similar to Equation (\ref{eq:WKB_Unstable}), indicating that the second term on the left side of Equation (\ref{eq:WKB_Unstable}) corresponds to compressibility effects.
The reason for the appearance of the product of  $\kappa$ and $\nu$ will be explained in Section VI B.

\subsection{Criterion of the stability}
\label{subsec:sigma0}
In this section, we analyze the system without using the WKB approximation to confirm that the stability criterion is determined by values such as the Rayleigh number. 
By substituting $\sigma = 0$ and combining Equations (\ref{eq:linear_EoC}) - (\ref{eq:linear_EoS}), we reduce the problem to a sixth-order differential equation for $\delta T$ alone. The resulting equation is:
\begin{align}
\begin{aligned}
\sum_{i=0}^{6} a_i \dfrac{d^{(i)} \delta T}{dz^{(i)}} = 0,
\end{aligned}
\label{eq:6thODEs_dT}
\end{align}
where $a_6 = 1$ and $a_i\ (i = 1,2,3,4,5)$ are functions of $k_{x,{\rm crit}}$, $1/H_T$, and $1/H_P \equiv \mu g/RT$. Only $a_0$ includes the diffusion coefficients, which is analogous to the Rayleigh number. The detailed notations are as follows:
\begin{align}
& a_5 = \dfrac{1}{H_P} - \dfrac{1}{H_T},
\label{eq:6thODEs_a5}\\[2mm]
& a_4 = 
3k_{x,{\rm crit}}^2 + \dfrac{4}{H_T} \left ( \dfrac{1}{H_P} - \dfrac{1}{H_T} \right ),
\label{eq:6thODEs_a4}\\[2mm]
& a_3 = 
-\left ( \dfrac{2}{H_P} - \dfrac{4}{H_T} \right ) k_{x,{\rm crit}}^2
+ \dfrac{8}{H_T^2} \left ( \dfrac{1}{H_P} - \dfrac{1}{H_T} \right ),
\label{eq:6thODEs_a3}\\[2mm]
& a_2 = 
k_{x,{\rm crit}}^4
- \dfrac{6}{H_T} \left ( \dfrac{1}{H_P} - \dfrac{1}{H_T} \right )k_{x,{\rm crit}}^2
+ \dfrac{8}{H_T^3} \left ( \dfrac{1}{H_P} - \dfrac{1}{H_T} \right ),
\label{eq:6thODEs_a2}\\[2mm]
& a_1 = 
\left ( \dfrac{1}{H_P} - \dfrac{2}{H_T} \right ) k_{x,{\rm crit}}^4
- \dfrac{8}{H_T^2} \left ( \dfrac{1}{H_P} - \dfrac{1}{H_T} \right )
k_{x,{\rm crit}}^2 ,
\label{eq:6thODEs_a1}\\[2mm]
& a_0 = 
k_{x,{\rm crit}}^6
- \dfrac{8}{H_T^3} \left ( \dfrac{1}{H_P} - \dfrac{1}{H_T} \right ) 
k_{x,{\rm crit}}^2
- \dfrac{g\beta}{T\kappa \nu} k_{x,{\rm crit}}^2.
\label{eq:6thODEs_a0}
\end{align}
where $k_{x,{\rm crit}}$ represents the critical wave number corresponding to $\sigma = 0$. 
The above equation implies that the stability criterion for convection is determined by $1/H_P$, $1/H_T$, and a parameter analogous to the Rayleigh number, Equation (\ref{eq:rayleigh_number}).

\section{Numerical Aprroach}
\label{sec:numerical}
In this section, we numerically solve the original equations by using the shooting method and the numerical simulation. 
We perform the multiple analyses step by step.

\subsection{Shooting method}
\label{subsec:shooting}
In this section, we explain the shooting method to find the accurate dispersion relations.
To apply this method, we assume the perturbation takes the form $\delta f(t, z) = \delta f(z) \exp(\sigma t)$, where $\sigma$ is the growth rate.
First, we omit the second derivative terms for the $z$-direction in the equations of motion for simplicity, which corresponds  to the very small limit of viscosity.
We find that the inclusion of the second derivative term in the shooting method does not work. We suppose that it is due to the stiffness of the perturbation equations for the numerical integration in the $z$-direction.
Based on Equations (\ref{eq:linear_EoC})-(\ref{eq:linear_EoS}), we derive the following three ODEs:
\begin{align}
&
\dfrac{d\delta v_z}{dz} = 
-k_x \delta v_x^\prime - \dfrac{1}{\rho_0}\dfrac{d\rho_0}{dz}\delta v_z
-\dfrac{\sigma}{\rho_0}\delta \rho,
\label{eq:linearODE_EoC}\\[3mm]
&
\dfrac{d\delta P}{dz} =
-(\sigma + \nu k_x^2 ) \rho_0 \delta v_z - g\delta \rho,
\label{eq:linearODE_EoMz}\\[3mm]
&
\dfrac{d^2 \delta T}{dz^2} =
-\dfrac{\beta}{\kappa} \delta v_z + \dfrac{\sigma + \kappa k_x^2}{\kappa} \delta T,
\label{eq:linearODE_Energy}\\[3mm]
&
\delta v_x^\prime = \dfrac{k_x}{\sigma + \nu k_x^2} \delta P,
\label{eq:linearODE_EoMx}\\[3mm]
&
\delta \rho = 
\rho_0 \left ( \dfrac{\delta P}{P_0} - \dfrac{\delta T}{T_0} \right ),
\label{eq:linearODE_EoS}
\end{align}
where $i\delta v_x$ is replaced by a new variable, $\delta v_x^\prime$.
The correction of growth rates to account for viscous effects expressed as second-order derivatives of $z$ will be discussed in Section \ref{subsubsec:correction}.
We numerically integrate the above ODEs from $z = 0$ toward $z = L_z$ using the fourth-order Runge-Kutta integrator and seek the growth rate $\sigma$ utilizing the shooting
method.

\subsubsection{Boundary conditions}
\label{subsubsec:boundary}
We can determine the growth rate of the ODEs (\ref{eq:linearODE_EoC}) - (\ref{eq:linearODE_EoS}) applying four appropriate boundary conditions\cite{TOBIAS199843}. 
We adopt the following boundary conditions:
\begin{align}
&
\delta T (0) = 0,
\label{eq:linear_boundary_bottom_vx}\\[1mm]
&
\delta v_z (0) = 0,
\label{eq:linear_boundary_bottom_vz}\\[1mm]
&
\delta T (L_z) = 0,
\label{eq:linear_boundary_top_vx}\\[1mm]
&
\delta v_z (L_z) = 0.
\label{eq:linear_boundary_top_vz}
\end{align}
These conditions represent isothermal rigid walls at both the bottom and top boundaries. 
The boundary conditions for the $z$-component of velocity correspond to to our choice of rigid wall boundary condition at $z=0$ and $L_z$. 

To analyze a different situation without viscosity we also consider the isothermal free boundary condition:
\begin{align}
&
\delta T (0) = 0,
\label{eq:linear_boundary_bottom_free_T}\\[3mm]
&
\delta P (0) + \dfrac{\delta v_z (0)}{\sigma}\left ( \dfrac{dP_0}{dz}\right )_{z = 0} = 0,
\label{eq:linear_boundary_bottom_free_P}\\[3mm]
&
\delta T (L_z) = 0,
\label{eq:linear_boundary_top_free_T}\\[3mm]
&
\delta P (L_z) + \dfrac{\delta v_z (L_z)}{\sigma}\left ( \dfrac{dP_0}{dz}\right )_{z = L_z} = 0.
\label{eq:linear_boundary_top_free_P}
\end{align}
Equations \eqref{eq:linear_boundary_bottom_free_P} and \eqref{eq:linear_boundary_top_free_P} require Lagrangian perturbations of pressure vanish at the deformed boundaries.
This boundary condition is also expected to be a good approximation for the case where the computational domain is sandwiched by the fluid that is light and stable for convection.

\subsubsection{Corrections of growth rate}
\label{subsubsec:correction}
In this study, we have prepared to solve the dispersion relation while neglecting the second derivative terms in the $z$-direction in the equations of motion. 
However, it is important to note that viscosity significantly impacts convective instability, as mentioned in Section \ref{sec:analytical}. 
Therefore, we correct the growth rates obtained in the previous analysis to account for these effects.

By using the equation of motion in the $z$-direction (\ref{eq:linear_EoMz}) and utilizing the previously determined growth rate, we modify the growth rate as follows:
\begin{align}
\sigma_{\rm cor} = \sigma + \dfrac{\nu}{\delta v_z}\dfrac{d^2 \delta v_z}{dz^2}.
\label{eq:eigenvalue_correction}
\end{align}
where $\sigma_{\rm cor}$ denotes the corrected growth rate. 
The second-order derivative of the eigenfunction, $d^2 \delta v_z / dz^2$, is obtained by finite deference approximation using the eigenfunction obtained by the shooting method.
Additionally, the correction terms are spatially averaged in the $z$-direction. 
This correction ensures consistency with the equation of motion in the $z$-direction, although not in the $x$-direction. 
The quantitative examination of this correction is presented in Section \ref{subsec:wViscosity}.

\subsection{Linear simulation of the time-dependent equations}
\label{subsec:simulation}
In this section, we derive the dispersion relations in the presence of both thermal conduction and viscosity by performing time-dependent numerical simulations. 
By this method we can examine the approximation invoked in Section \ref{subsec:shooting}.

We directly solve Equation (\ref{eq:linear_EoC})--(\ref{eq:linear_EoS}).
The simulations utilize an Eulerian grid scheme, where the domain height is discretized into 128 grid points. 
The boundary conditions at the bottom and top boundaries are set as isothermal rigid walls, defined as:
\begin{align}
    &
    \delta T(0) = 0,
    \label{eq:simu_boundary_bottom_T}\\[1mm]
    &
    \delta v_x^\prime(0) = 0,
    \label{eq:simu_boundary_bottom_vx}\\[1mm]
    &
    \delta v_z(0) = 0,
    \label{eq:simu_boundary_bottom_vz}\\[1mm]
    &
    \delta T(L_z) = 0,
    \label{eq:simu_boundary_top_T}\\[1mm]
    &
    \delta v_x^\prime(L_z) = 0,
    \label{eq:simu_boundary_top_vx}\\[1mm]
    &
    \delta v_z(L_z) = 0.
    \label{eq:simu_boundary_top_vz}
\end{align}

The initial perturbation is introduced as $\delta T(t = 0, z) = \sin(\pi z)$, with its initial amplitude set to 1. 
The growth rate is then defined based on the amplitude of the temperature perturbation, $\delta T(t, z)$, at time $t$, after the perturbation has grown significantly.
The growth rate is calculated using the following definition:
\begin{align}
\sigma \equiv \dfrac{1}{\Delta t} \log \left\{ \dfrac{\mathrm{Amp}[\delta T(t+\Delta t, z)]}{\mathrm{Amp}[\delta T(t, z)]} \right\},
\label{eq:simu_growthrate}
\end{align}
where $\text{Amp}[\delta T (t, z)]$ represents the amplitude of the temperature fluctuation at time $t$.

\section{Results of numerical analyses}
\label{sec:results}
This section presents the results of numerical analyses step by step.
First, Section \ref{subsec:structure} provides an overview of the flow structure for the case where $\kappa / \sqrt{gL_z^3} = 10^{-2.0}$ and $\nu / \sqrt{gL_z^3} = 0.0$.
This is for the verification of our analyses.
Then, we examine the case without dissipation in Section \ref{subsec:woDispation}, focusing on convection dynamics absent of any dissipative effects.
Next, Section \ref{subsec:woViscosity} explores the case of thermal conduction without viscosity.
Section \ref{subsec:woTC} is for the case of viscosity without thermal conduction and Section \ref{subsec:wViscosity} is for the case including both thermal conduction and viscosity.
Finally, we discuss the critical wavelength in Section \ref{subsec:critical}.

\subsection{Flow structures}
\label{subsec:structure}
\begin{figure}[h]
\centering
\includegraphics[width=\linewidth]{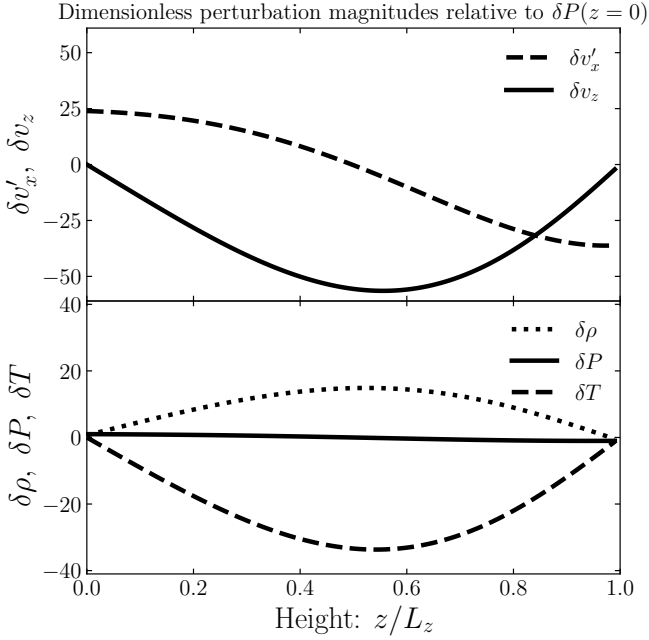}
\caption{Vertical profiles of the perturbed physical values in the case of $H_T / L_z = 2$, $\kappa / \sqrt{gL_z^3} = 10^{-2.0}$ and $\nu / \sqrt{gL_z^3} = 0$: $\delta v_x^\prime, \delta v_z$ in the top panel, $\delta \rho, \delta P, \delta T$ in the bottom panel. The plotted values are relative to the pressure perturbation at the bottom boundary $\delta P (0)$, the gravitational acceleration $g$, the system height $L_z$, and the mid-plane temperature $T_{\rm mid}$.
}
\label{fig:Eigenfuc_Amp}
\end{figure}

In this section, we initially present the eigenfunctions to prove that our analysis is capable of accurately capturing the convective motion. 
Figure \ref{fig:Eigenfuc_Amp} illustrates the amplitudes of the eigenfunctions for the most unstable wave number $k_{x,{\rm max}}$ in the case where $H_T / L_z = 2$, $\kappa / \sqrt{gL_z^3} = 10^{-2.0}$ and $\nu / \sqrt{gL_z^3} = 0$.
The figure shows that the convective structure is not perfectly symmetric with respect to the mid-plane.
This is a consequence of the inhomogeneity of the unperturbed state.
In contrast, for the case of $H_T/L_z = 100$, note that the solutions are nearly symmetric sinusoidal and cosinusoidal functions with respect to $z/L_z = 0.5$.
Additionally, the amplitude of perturbed pressure, $\delta P/P_0$, is found to be an order of magnitude smaller than the density, $\delta \rho /\rho_0$,  and temperature perturbation, $\delta T / T_0$.
This confirms that convection in this system is a sub-sonic phenomenon.

\begin{figure*}[ht]
\centering
\includegraphics[width=0.9\linewidth]{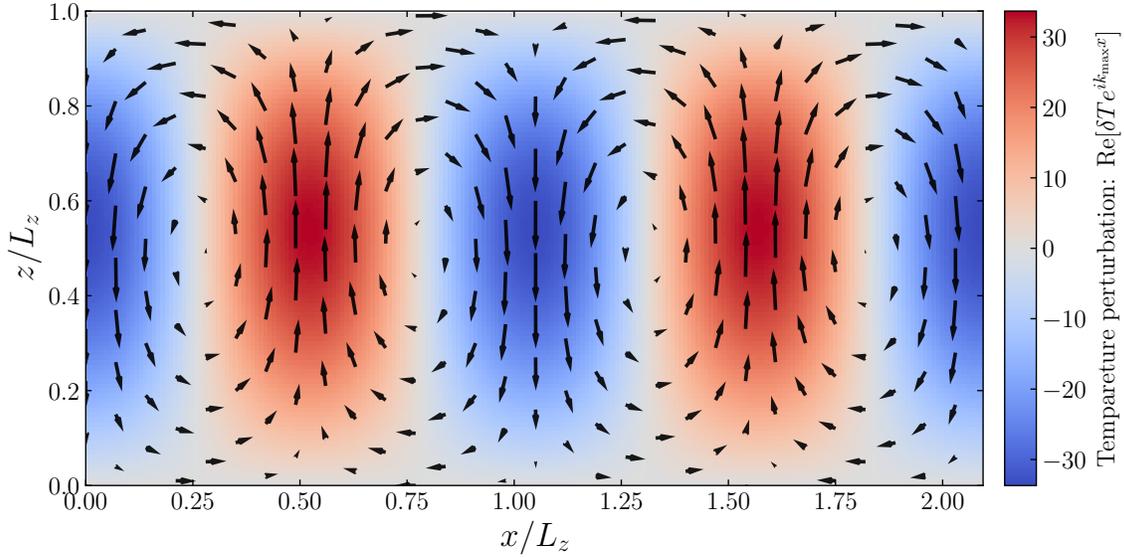}
\caption{
Structure and motion of gas in the case of $H_T / L_z = 2$, $\kappa / \sqrt{gL_z^3} = 10^{-2.0}$ and $\nu / \sqrt{gL_z^3} = 0$. 
The color represents $\mathrm{Re}[\delta T e^{ik_x x}]$, and the arrows denote $(\mathrm{Re}[\delta v_x^\prime e^{ik_x x}], \mathrm{Re}[\delta v_z e^{ik_x x}])$, respectively. 
All values are normalized by the thermal flux perturbation at the bottom boundary.
}
\label{fig:Eigenfunc_structure}
\end{figure*}

Figure \ref{fig:Eigenfunc_structure} provides a complementary illustration of this phenomenon, showing two-dimensional maps of the gas temperature perturbations and their corresponding velocity perturbations for the same $k_x$. 
The analysis reveals that the gas in the upper layer exhibits a higher velocity than that in the lower layer, a result of not neglecting the vertical structure in the unperturbed state.

\subsection{Simple cases without dissipation}
\label{subsec:woDispation}
In this section, we present the results for the case excluding all dissipation. 
In this case, since the imaginary part of the dispersion relation (\ref{eq:WKB_DR_Im}) vanishes, the analytical solution can be derived without any difficulty: from Equations (\ref{eq:WKB_DR_kz0}) and (\ref{eq:WKB_DR_boussinessq}), which were obtained using the WKB approximation:
\begin{align}
\sigma^2 &= \dfrac{1}{2} \left [
-\left ( \dfrac{RT}{\mu} k^2 + \dfrac{g}{H_\rho} \right )
+ \sqrt{
\left ( \dfrac{RT}{\mu} k^2 + \dfrac{g}{H_\rho} \right )^2
+4\dfrac{gR\beta}{\mu} k_x^2
}
\right ],
\label{eq:woDispation_WKB_DR}
\\
\sigma^2 &= \dfrac{g}{H_T} \dfrac{k_x^2}{k^2}.
\label{eq:woDispation_WKB_DR_boussinessq}
\end{align}
Then, the dispersion relation for the fundamental mode can be obtained by substituting $k_zL_z = \pi$ into the aforementioned equations.
In addition, this mode that rotates the entire vertical direction of the domain with $k_z L_z = \pi$ is the most unstable.

\begin{figure}[h]
\centering
\includegraphics[width=\linewidth]{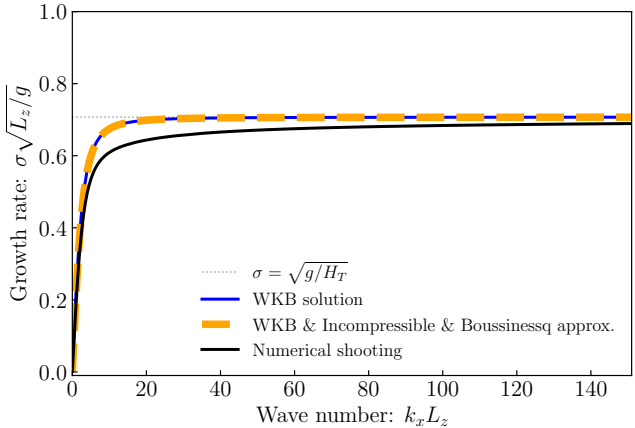}
\caption{Dispersion relation of the fundamental mode ($k_z L_z = \pi$) in the case without any dissipation and $H_T / L_z = 2$.
The blue solid line is the WKB solution derived from Equation (\ref{eq:woDispation_WKB_DR}), while the orange dashed line, derived from Equation (\ref{eq:woDispation_WKB_DR_boussinessq}), applying incompressible and Boussinesq approximation additionally.
The black solid line is the numerical solution by using shooting method that is regarded as the correct solution, and gray dashed line corresponds to $\sigma = \sqrt{g/H_T}$.}
\label{fig:HT2_DR_Nodisp}
\end{figure}

\begin{figure}[h]
\centering
\includegraphics[width=\linewidth]{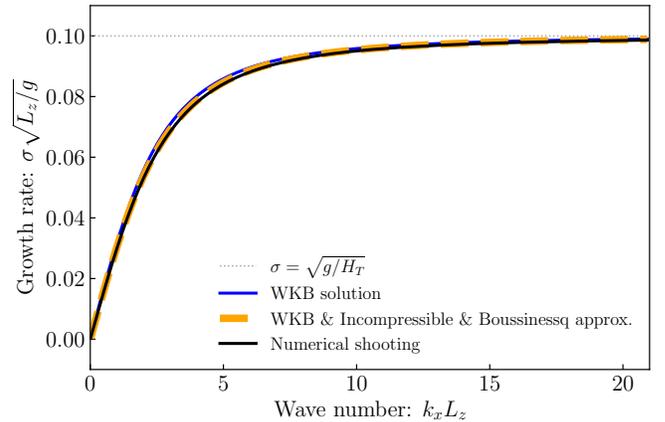}
\caption{Dispersion relation of the fundamental mode ($k_z L_z = \pi$) in the case without any dissipation and $H_T / L_z = 100$.
}
\label{fig:HT100_DR_Nodisp}
\end{figure}

Figures \ref{fig:HT2_DR_Nodisp} and \ref{fig:HT100_DR_Nodisp} show the dispersion relations for the cases of $H_T / L_z = 2$ and $100$. 
In the case of $H_T / L_z = 2$, the WKB solution deviates from the numerical solution, while in the case of $H_T / L_z = 100$, they are almost identical.
This is because the sinusoidal functions adopted in WKB approximation is not close to the correct eigenfunction in the case of significantly varying vertical structure.
Conversely, in the case of $H_T/L_z=100$, since the temperature scale height is much larger than the system height, the change in the unperturbed state in the vertical direction is sufficiently small to allow the sinusoidal functions to be good approximations for the eigenfunctions.

The common features of the results in Figures \ref{fig:HT2_DR_Nodisp} and \ref{fig:HT100_DR_Nodisp} are that the growth rate increases with increasing wave number, and both the analytical and numerical solutions converge to the asymptotic behaviour. 
In particular, they approach the value $\sigma \to \sqrt{H_T / g}$.
This result implies that the growth rate is characterized by the free fall time over the distance of the temperature scale height.

\subsection{Effect of thermal conduction without viscosity}
\label{subsec:woViscosity}
\begin{figure}[h]
\centering
\includegraphics[width=\linewidth]{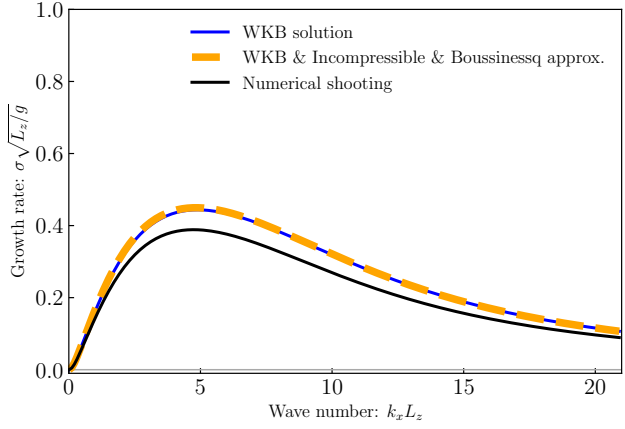}
\caption{Dispersion relation of the fundamental mode ($k_z L_z = \pi$) in the case of including thermal conduction without viscosity for $H_T / L_z = 2$, setting $\kappa / \sqrt{gL_z^3} = 10^{-2.0}$ and $\nu / \sqrt{gL_z^3} = 0.0$.}
\label{fig:H2_DR_woAllViscosity}
\end{figure}

\begin{figure}[h]
\centering
\includegraphics[width=\linewidth]{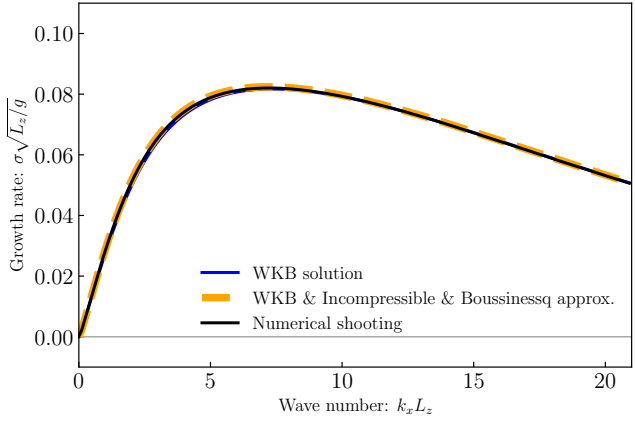}
\caption{Dispersion relation in the case of including thermal conduction without viscosity for $H_T / L_z = 100$, setting $\kappa / \sqrt{gL_z^3} = 10^{-3.5}$ and $\nu / \sqrt{gL_z^3} = 0.0$.}
\label{fig:H100_DR_woAllViscosity}
\end{figure}

We then derive the dispersion relation in the case of finite thermal conduction and vanishing viscosity in this section. 
In this case, we obtain the WKB solution by solving the quartic equation (\ref{eq:WKB_DR_kz0}) and the quadratic equation (\ref{eq:WKB_DR_boussinessq}). 
Figures \ref{fig:H2_DR_woAllViscosity} and \ref{fig:H100_DR_woAllViscosity} show the dispersion relations in the case where viscosity is excluded but thermal conduction is included for $H_T / L_z = 2$ and $100$. 
We set the normalized thermal diffusivity as $\kappa / \sqrt{gL_z^3} = 10^{-2.0}$ for $H_T / L_z = 2$ and $\kappa / \sqrt{gL_z^3} = 10^{-3.5}$ for $H_T / L_z = 100$.

As illustrated in Figure \ref{fig:H2_DR_woAllViscosity}, when the vertical variation within the domain is significant, a discrepancy arises between the WKB solution and the exact numerically determined solution. 
Conversely, in Figure \ref{fig:H100_DR_woAllViscosity}, where the vertical variation is sufficiently small, the WKB solution and the numerical solution show a high degree of agreement. 
Notably, on the short wavelength side, the analytical solution asymptotically approaches the numerical solution.

As shown in Figures \ref{fig:H2_DR_woAllViscosity} and \ref{fig:H100_DR_woAllViscosity}, the inclusion of the diffusion term in the model leads to a notable peak in the growth rate, which highlights the effect of thermal conduction on the convective instability. 
However, the convective instability cannot be stabilized by considering thermal conduction alone, so the unstable modes observed in Figures \ref{fig:HT2_DR_Nodisp} and \ref{fig:HT100_DR_Nodisp} remain unstable even when thermal conduction is taken into account.

\subsection{Effect of viscosity without thermal conduction}
\label{subsec:woTC}
\begin{figure}[h]
\centering
\includegraphics[width=\linewidth]{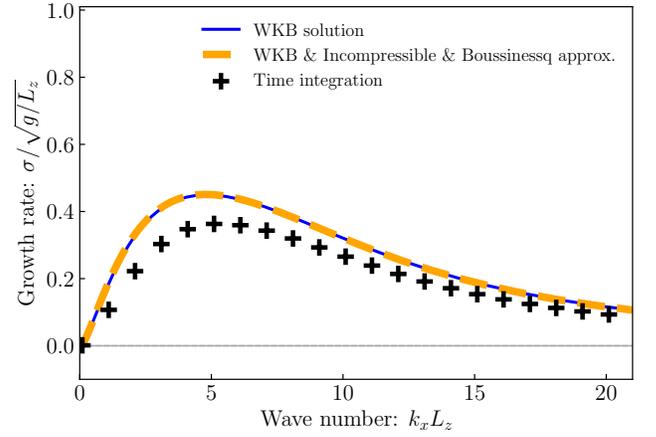}
\caption{Dispersion relation of the fundamental mode ($k_z L_z = \pi$) in the case of including viscosity without thermal conduction for $H_T / L_z = 2$, setting $\kappa / \sqrt{gL_z^3} = 0.0$ and $\nu / \sqrt{gL_z^3} = 10^{-2.0}$.
The plus symbols represent the growth rate obtained from the time integration.}
\label{fig:H2_DR_woAllTC}
\end{figure}

\begin{figure}[h]
\centering
\includegraphics[width=\linewidth]{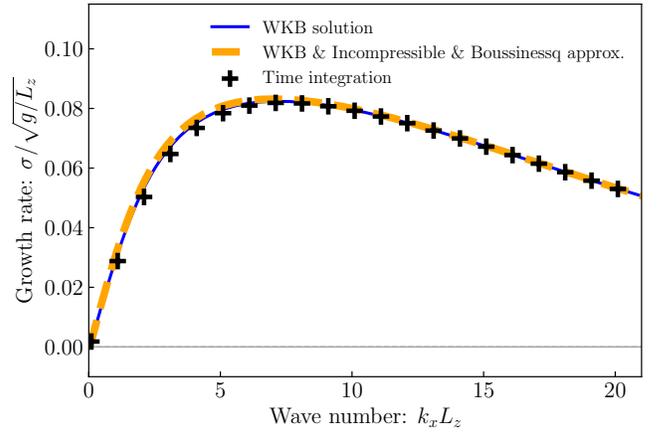}
\caption{Dispersion relation in the case of including viscosity without thermal conduction for $H_T / L_z = 100$, setting $\kappa / \sqrt{gL_z^3} = 0.0$ and $\nu / \sqrt{gL_z^3} = 10^{-3.5}$.}
\label{fig:H100_DR_woAllTC}
\end{figure}

We derive the dispersion relation in the case of finite viscosity and vanishing thermal conduction in this section.
In this case, we obtain the WKB solution by solving the quartic equation (\ref{eq:WKB_DR_kz0}) and the quadratic equation (\ref{eq:WKB_DR_boussinessq}). 
Figures \ref{fig:H2_DR_woAllTC} and \ref{fig:H100_DR_woAllTC} show the dispersion relations in the case where thermal conduction is excluded, but viscosity is included for $H_T / L_z = 2$ and $100$.
Plus symbols in these figures correspond to the simulation result and can be regarded as the exact solution.

As same as Section \ref{subsec:woViscosity}, the inclusion of viscosity in the analysis leads to a notable peak in the growth rate, the convective instability cannot be stabilized by considering viscosity alone.
Thus, the unstable modes observed in Figures \ref{fig:HT2_DR_Nodisp} and \ref{fig:HT100_DR_Nodisp} remain unstable when only thermal conduction or viscosity is taken into account.

\subsection{Effect of combination of thermal conduction and viscosity}
\label{subsec:wViscosity}
\begin{figure}[h]
\centering
\includegraphics[width=1.0\linewidth]{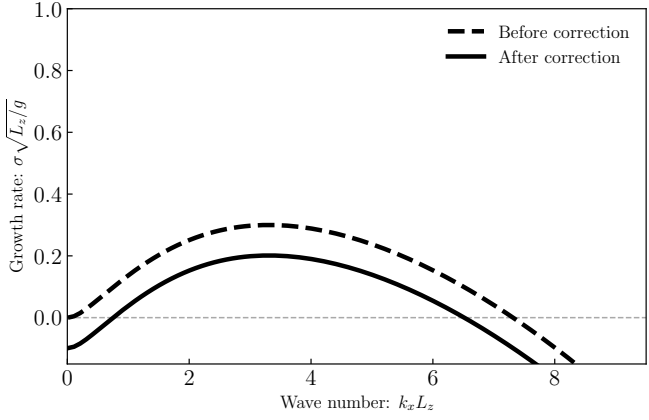}
\caption{The change of dispersion relation of the fundamental mode ($k_z L_z = \pi$) in the case of $H_T / L_z = 2$ and $\kappa / \sqrt{gL_z^3} = \nu / \sqrt{gL_z^3} = 10^{-2.0}$.
The dashed line is a result before correction, while solid line is after correction by utilizing Equation (\ref{eq:eigenvalue_correction}).}
\label{fig:corrected_DR}
\end{figure}

In this section, we introduce viscosity in addition.
As mentioned in Section \ref{sec:numerical}, solving the ODEs with viscosity presents mathematical challenges. 
Figure \ref{fig:corrected_DR} plots the corrected growth rate as a function of $k_x$ in the case of $H_T / L_z = 2$ and $\kappa / \sqrt{g L_z^3} = \nu / \sqrt{g L_z^3} = 10^{-2.0}$.
The correction is applied using Equation (\ref{eq:eigenvalue_correction}), resulting in the solid line. 
This correction demonstrates a decrease in the growth rate and stabilizes on the long wavelength side, highlighting the role of viscosity in suppressing instability within the system.
This behavior may highlight the stabilizing influence of viscosity, expressed as the second order derivative of $z$, especially in convection cells with a large aspect ratio.

\begin{figure}[h]
\centering
\includegraphics[width=1.0\linewidth]{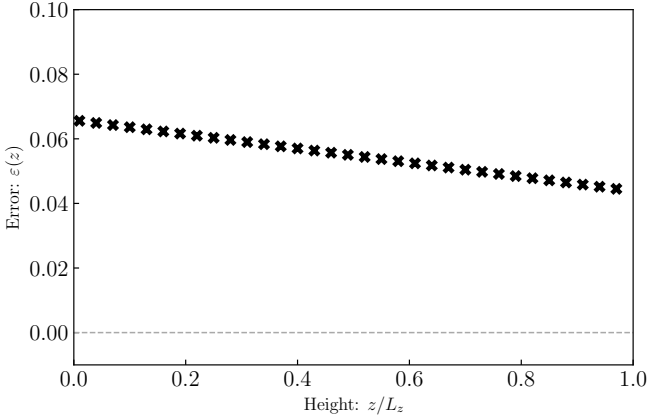}
\caption{The error profile defined in Equation (\ref{eq:error_verify}) and normalized by the eigenfunction of velocity in the horizontal direction.}
\label{fig:Varification_Correction}
\end{figure}

Figure \ref{fig:Varification_Correction} illustrates the verification results obtained by substituting the corrected growth rate from Equation (\ref{eq:eigenvalue_correction}) and the corresponding eigenfunctions into the equation of motion in the $x$-direction, given by Equation (\ref{eq:linear_EoMx}).
The profile of $\epsilon(z)$ is derived using the following equation:
\begin{align}
\begin{aligned}
    &
\epsilon (z) \sigma_{\rm cor} \delta v_x^\prime (z) = 
\\[1mm]
&
\sigma_{\rm cor}\delta v_x^\prime (z) - \dfrac{k_x}{\rho_0}\delta P (z) + \nu k_x^2 \delta v_x^\prime (z) - \nu \dfrac{d^2 \delta v_x^\prime (z)}{dz^2},
\end{aligned}
\label{eq:error_verify}
\end{align}
where $\epsilon (z)$ quantifies the difference between the left and right sides of the $x$-component of the momentum equation.
If $\epsilon (z)$ is small, the growth rate correction method is verified to be accurate.
In addition, We compute $d^2 \delta v_x^\prime / dz^2$ in the same way to obtain $d^2 \delta v_z / dz^2$.
The results show that the error throughout the system is only a few percent of the expected value.
This suggests that the viscous term expressed in the second derivative of $z$ may play a role in stabilizing the convection.
Therefore, we discuss the case where both dissipation terms are considered by using this correction method.

\begin{figure}[h]
\centering
\includegraphics[width=1.0\linewidth]{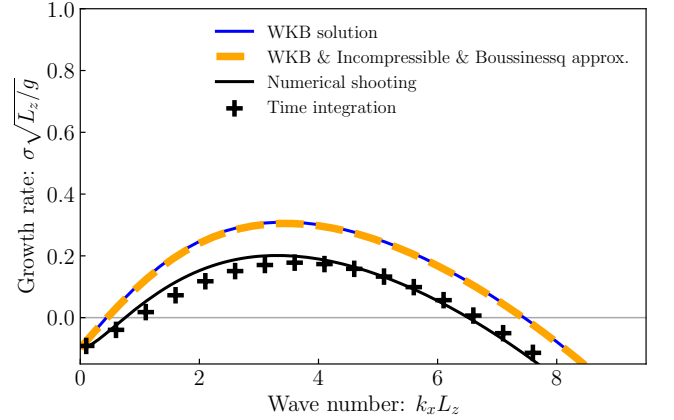}
\caption{Dispersion relation of the fundamental mode ($k_z L_z = \pi$) in the case of $H_T / L_z = 2$ and $\kappa / \sqrt{gL_z^3} = \nu / \sqrt{gL_z^3} = 10^{-2.0}$.}
\label{fig:H2_DR_wViscosity}
\end{figure}

\begin{figure}[h]
\centering
\includegraphics[width=1.0\linewidth]{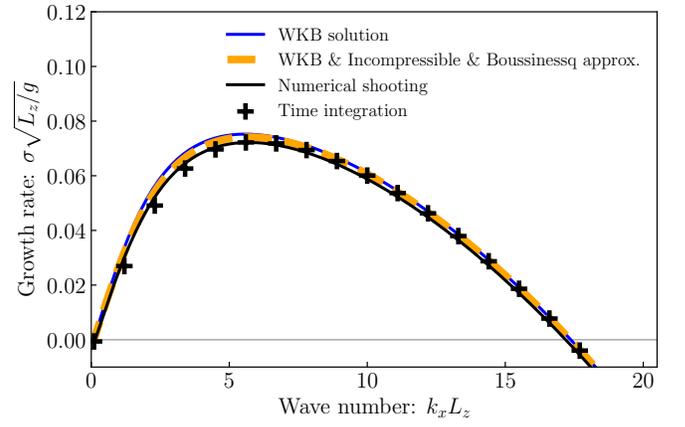}
\caption{Dispersion relation of the fundamental mode ($k_z L_z = \pi$) in the case of $H_T / L_z = 100$ and $\kappa / \sqrt{gL_z^3} = \nu / \sqrt{gL_z^3} = 10^{-3.5}$.}
\label{fig:H100_DR_wViscosity}
\end{figure}

Then, we compare the numerical solution with the WKB solution as in Sections \ref{subsec:woDispation} and \ref{subsec:woViscosity}.
Figures \ref{fig:H2_DR_wViscosity} and \ref{fig:H100_DR_wViscosity} show the results for the cases of $H_T / L_z = 2$ and $100$.
As in Sections \ref{subsec:woDispation} and \ref{subsec:woViscosity}, the WKB solution approaches the numerical solution on the short wavelength side when the temperature scale height is much larger than the domain height.
However, in contrast to Sections \ref{subsec:woDispation} and \ref{subsec:woViscosity}, we observe the emergence of wavelengths where instability is suppressed by diffusion effects.
We also emphasize that our correction to the growth rate is valid, as the corrected dispersion relation (black solid line) is nearly identical to the exact solution (triangular scatter points) obtained from numerical simulations.

Moreover, especially in this case, the growth rate obtained from the WKB approximation is several orders of magnitude larger than that of the numerical solution, leading to an overestimation of the unstable wavelength range.

\subsection{Determining factors for critical wavelengths}
\label{subsec:critical}
\begin{figure}[h]
\centering
\includegraphics[width=\linewidth]{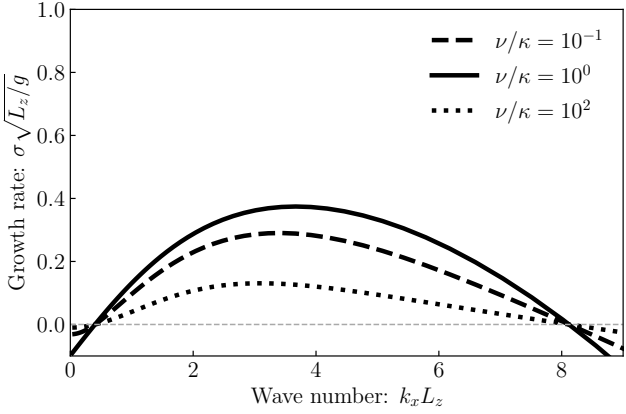}
\caption{Dispersion relation for the case of $H_T / L_z = 2$ and $g L_z^3/ \kappa \nu = 10^4$.
The difference of each lines is from the ratio of diffusion coefficients $\nu / \kappa$.
The black dashed line is $\nu/\kappa = 10^{-1}$, the red solid  line is $\nu/\kappa = 10^{0}$ and the blue dotted line is $\nu/\kappa = 10^{2}$.
The gray dashed line corresponds to $\sigma = 0$.
}
\label{fig:DR_woVisc_Ra5}
\end{figure}

In this section, we focus on verifying the stability analysis by examining the critical wave number where $\sigma = 0$. 
Figure \ref{fig:DR_woVisc_Ra5} shows the dispersion relations obtained by solving Equation (\ref{eq:WKB_DR_kz0}), for a fixed value of $H_T / L_z = 2$ and $g L_z^3/\kappa \nu = 10^4$, with variations of the ratio $\nu / \kappa$ set to $10^{-1}$, $10^{0}$, and $10^{2}$. This variation shows how the dispersion relation changes with different ratios of viscosity to thermal diffusivity.

While the specific details of the dispersion relation $\sigma \neq 0$ differ depending on the ratio, the critical wave number where $\sigma = 0$ remains consistent across the different cases. 
This consistency confirms the results derived from Equation (\ref{eq:6thODEs_dT}), which indicates that the critical wavenumber is not affected by the changes in the ratio of $\nu$ to $\kappa$.
In other words, the product of the dissipation coefficients $gL_z^3 / \kappa\nu$, which is similar to Rayleigh number $g\beta L_z^4 / T\kappa\nu$, plays a crucial role in determining the critical wavelength.

\section{Discussions}
\label{sec:discussion}
In this section, we carefully examine the results obtained in Section \ref{sec:results}. 
First, we discuss the contribution of the diffusion terms by showing the change of the dispersion relation in Sections \ref{subsec:contribution_diffusion}.
Then, in Section \ref{subsec:supriority}, we give the quantitative understanding to explain why both of diffusion terms are necessary for stabilization, by re-focusing the constant term of the analytical dispersion relation (\ref{eq:WKB_DR_kz0}).
In Section \ref{subsec:HT_Racrit_relation}, we also propose the critical Rayleigh number as a function of temperature scale height.
Finally, Section \ref{subsec:comparison_Boussinessq} verifies the Boussinesq approximation using the shooting method.

\subsection{Contribution of diffusion terms}
\label{subsec:contribution_diffusion}

\begin{figure}[htbp]
\centering
\includegraphics[width=0.85\linewidth]{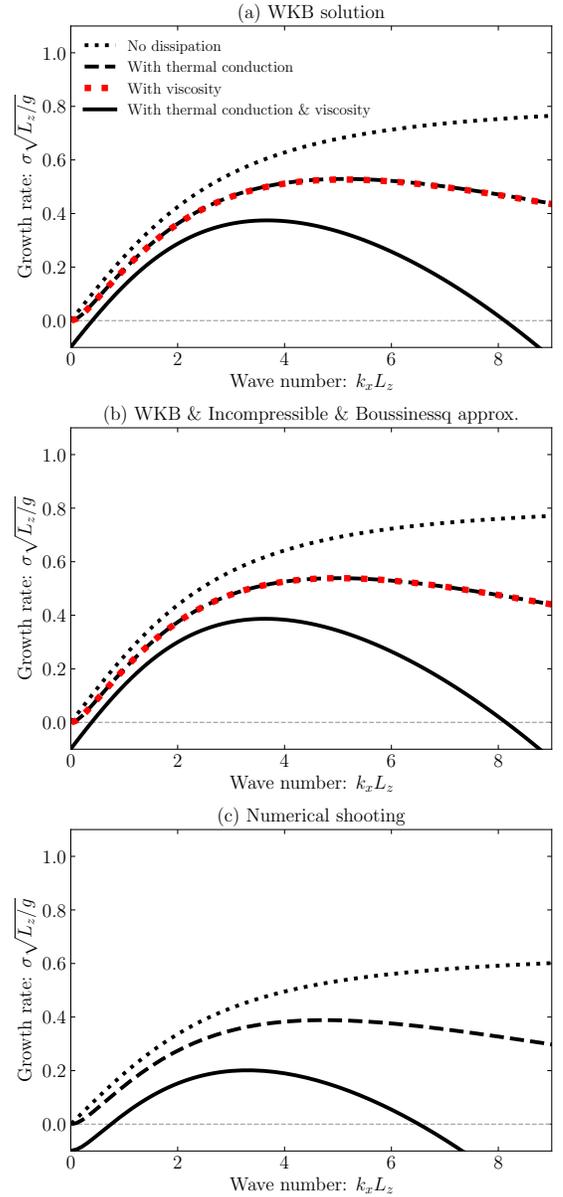}
\caption{Changes in the dispersion relations when thermal and viscous diffusion are considered. Figure (a) displays the WKB solution derived from Equation (\ref{eq:WKB_DR_kz0}), while Figure (b) incorporates the incompressible and Boussinesq approximations, as shown in Equation (\ref{eq:WKB_DR_boussinessq}). Figure (c) presents the numerical solution obtained using the shooting method. The black dotted, dashed, and solid lines represent the cases of no dissipation, inclusion of thermal conduction, and additional viscosity, respectively. Additionally, Figures (a) and (b) include the case considering only viscosity, indicated by the red dashed line.
}
\label{fig:discuss_change_DR}
\end{figure}

In this section, we analyze the role of diffusion in modifying the stability of convection by comparing the dispersion relations with and without the diffusion terms. Figure \ref{fig:discuss_change_DR} shows the changes in the dispersion relations when both thermal and viscous diffusion are considered, specifically for the case where $H_T / L_z = 2$ and $\kappa / \sqrt{gL_z^3} = \nu / \sqrt{gL_z^3} = 10^{-2.0}$. 

Figure (a) shows the WKB solution derived from Equation (\ref{eq:WKB_DR_kz0}), (b) shows the result obtained by additionally applying the incompressible and Boussinesq approximation as shown in Equation (\ref{eq:WKB_DR_boussinessq}), while Figure (c) displays the numerical solution obtained using the shooting method. The black dotted, dashed, and solid lines represent the cases of no dissipation, considering thermal conduction, and additional viscosity, respectively. Additionally, Figures (a) and (b) also show the case where only viscosity is considered, represented by the red dashed line.

It seems that both the WKB solutions and the numerical solutions show a consistent trend: 
when only one type of diffusion is taken into account, the growth rate is reduced, but the instability is not completely suppressed. 
However, when both types of diffusion are considered, critical wavelengths emerge on both the long and short wavelength sides of the system. 
This indicates that both diffusions are necessary to stabilize convection.

\begin{figure}[t]
    \centering
    \includegraphics[width=\linewidth]{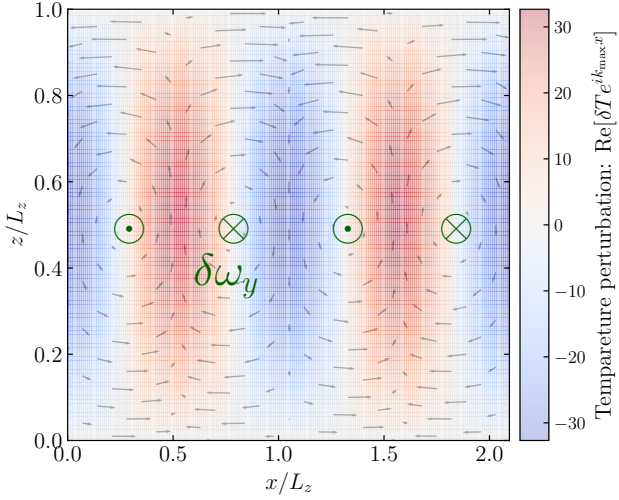}
    \caption{The vorticity mode $\delta \omega_y$ added to Figure \ref{fig:Eigenfunc_structure}, shown as green color symbols.}
    \label{fig:discuss_vortex_eigen}
\end{figure}

We must understand the reason why both thermal conduction and viscosity are required for stabilization. 
To analyze this, we introduce the vorticity in the $y$-direction, defined as $\delta \omega_y = ik_z \delta v_x - ik_x \delta v_z$.
Using this definition, we can transform the linearized equations under the incompressible and Boussinesq approximations as follows:
\begin{align}
    &
    \sigma \delta \Omega_y = \dfrac{g}{T}k_x \delta T - \nu k^2 \delta \Omega_y,
    \label{eq:discuss_vortex}\\[1mm]
    &
    \sigma \delta T = \beta \dfrac{k_x}{k^2}\delta \Omega_y - \kappa k^2 \delta T,
    \label{eq:discuss_temp}
\end{align}
where $\delta \Omega_y \equiv i \delta \omega_y$, so note that the phase of the true vorticity, $\delta \omega_y$, is shifted by $\pi/2$ compared with that of the temperature perturbation, $\delta T$.
Figure \ref{fig:discuss_vortex_eigen} shows the vorticity mode $\delta \omega_y$, added to Figure \ref{fig:Eigenfunc_structure} and displayed as green symbols. 
The figure indicates that the peaks of $\delta \omega_y$ are located where $\delta T = 0$.
The first equation (\ref{eq:discuss_vortex}) represents the evolution of the vorticity mode, while the second equation (\ref{eq:discuss_temp}) describes the evolution of the entropy mode. 
Equations (\ref{eq:discuss_vortex}) and (\ref{eq:discuss_temp}) demonstrate that these two modes are coupled by gravity, as they would be independent of each other in the absence of gravity (see Appendix \ref{sec:couple}).

First, let's consider the case of both $\kappa=0, \nu=0$.
In Equation (\ref{eq:discuss_temp}), the first term on the right-hand side represents the effect of advection on temperature transport. 
Considering the situation where a fluid element is going upward, the local temperature at fixed Eulerian coordinates increases because its entropy is greater than that of its surroundings, and the increased temperature results in a larger buoyancy and accelerates the upward going motion as shown in Equation (\ref{eq:discuss_vortex}).
Likewise, the fluid element is going downward, the temperature decreases and accelerate the downward motion.
This is the main cause of convection. 

Then, let us return to the discussion of why both dissipation processes are needed for stabilization. 
Let's consider perturbations of $\delta T > 0$ and $\delta \Omega_y > 0$. 
In this case, if viscosity vanishes, a perturbation with sufficiently small $\delta T / \delta \Omega_y$ make the right hand side of Equation (\ref{eq:discuss_temp}) to be positive, which means the growth of the perturbation.
Consequently, the entropy mode grows, which in turn drives the growth of the vortex mode, as described by Equation (\ref{eq:discuss_vortex}). 
A similar argument applies if thermal conduction is neglected: perturbations with sufficiently large $\delta T / \delta \Omega_y$ should grow.

\begin{figure}[t]
    \centering
    \includegraphics[width=0.9\linewidth]{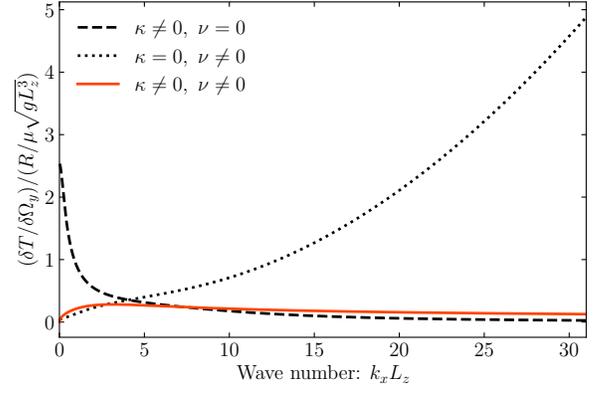}
    \caption{Normalized ratio of the amplitudes of $\delta \Omega_y$ and $\delta T$ as a function of the wave number $k_x$ for the case of $H_T / L_z = 2$. 
    The black dashed line represents the case where $\kappa \neq 0$ and $\nu = 0$, the black dotted line corresponds to $\kappa = 0$ and $\nu \neq 0$, and the red solid line represents the case where $\kappa \neq 0$ and $\nu \neq 0$. 
    When a diffusion coefficient, $D$, is non-zero, its value is set to $D / \sqrt{gL_z^3} = 10^{-2.0}$.
    }
    \label{fig:discuss_EigenRatio}
\end{figure}

Figure \ref{fig:discuss_EigenRatio} illustrates the normalized ratio of the amplitudes of $\delta \Omega_y$ and $\delta T$ as a function of the wave number, $k_x$.
In the case of $\kappa \neq 0$ and $\nu = 0$ (black dashed line), this ratio decreases with increasing wave number.
Therefore, even on the short-wavelength side, where the diffusion term $\kappa k^2$ becomes significant, the driving force $\beta \delta \Omega_y$ remains more dominant than the diffusion term $\kappa k^2 \delta T$ in Equation (\ref{eq:discuss_temp}).
This result agrees with the preceding explanation.
In contrast, for $\kappa = 0$ and $\nu \neq 0$ (black dotted line), the ratio increases with increasing wave number, and is consistent with expected behavior.
On the other hand, in the case of $\kappa \neq 0$ and $\nu \neq 0$ (solid red line), the ratio does not increase or decrease sufficiently. 
Because this ratio is neither large enough nor small enough to dominate, the driving forces $\delta \omega_y$ and $\delta T$ are stabilized by diffusion, resulting in the stabilization of the convective instability. 
Even in cases where the system becomes unstable, the growth rate of the instability is significantly reduced because of the combined effects of thermal conduction and viscosity. 
This suggests that while instability may, in principle, persist, its impact is likely to be negligible over observable timescales.  

\subsection{Competition between driving force and diffusion}
\label{subsec:supriority}
In this section, we re-examine the stability of convection in the context of the superiority of driving force and diffusion. We focus on the constant terms in the inequalities (\ref{eq:WKB_Unstable}) and (\ref{eq:WKB_Unstable_boussinessq}). Normalized by the domain height $L_z$, we obtain the following inequalities by utilizing the determination of Rayleigh number, Equation (\ref{eq:rayleigh_number}), which provides the criterion for convective instability:
\begin{align}
&(kL_z)^6 + \dfrac{\mu g L_z^2}{RT H_\rho} (kL_z)^4 - R{\rm a} (k_x L_z)^2 < 0,
\label{eq:WKB_Compressive_k-Ra}
\\[3mm]
&(kL_z)^6 - R{\rm a} (k_x L_z)^2 < 0.
\label{eq:WKB_Chandra_k-Ra}
\end{align}
These inequalities are consistent with the relationship between \(k_x\) and Rayleigh number mentioned in earlier work (e.g., Chandrasekhar 1981\cite{Chandrasekhar1981}):
\begin{align}
R{\rm a} > \dfrac{[\pi^2 +(k_x L_z)^2]^3}{(k_x L_z)^2}.
\label{eq:discuss_Chandra_k_Ra}
\end{align}
This indicates that the \(k_x - R{\rm a}\) relation reflects the superiority of driving force over diffusion, as shown in Section \ref{subsec:contribution_diffusion}. 
Conversely, convection is attenuated by diffusion, and the critical wavelength is determined by factors such as the Rayleigh number.

\begin{figure}[t]
\centering
\includegraphics[width=0.9\linewidth]{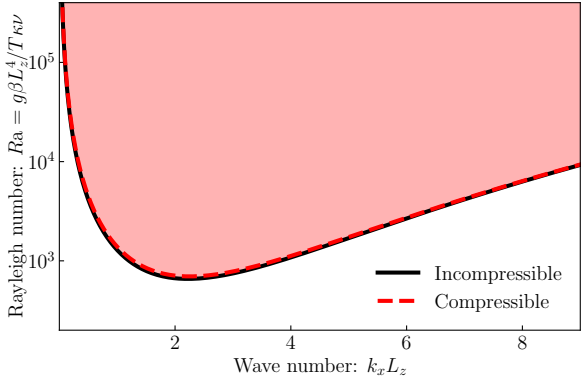}
\caption{
The relation between $k_x$ and Rayleigh number $R{\rm a}$ for the fundamental mode $(k_z L_z = \pi)$.
The wave number is normalized by the system height.
The red dashed and black solid lines represent the relationships derived from Equations (\ref{eq:WKB_Compressive_k-Ra}) and (\ref{eq:WKB_Chandra_k-Ra}).
}
\label{fig:kRa_relation}
\end{figure}

We then compare our analytical work with the previous work\cite{Chandrasekhar1981} that investigated the relationship between the horizontal wave number $k_x$ and the Rayleigh number in the context of convective instability. 
Figure \ref{fig:kRa_relation} illustrates one such relationship for a system for the fundamental mode $(k_z L_z = \pi)$ in the case of $H_T/L_z = 2$ by following Equations (\ref{eq:WKB_Compressive_k-Ra}) and (\ref{eq:WKB_Chandra_k-Ra}). 
The red shaded region is thought to represent the area identified by Chandrasekhar where convection is likely to occur.
Chandrasekhar's analysis suggests that convection can occur within this red-shaded region, indicating that certain conditions of $k_x$ and Rayleigh number are necessary for instability to grow. 
Comparing this with our analysis, we observe a consistent result indicating the stabilization of convection. 
While Chandrasekhar's theory postulates that long and short wavelengths are stable, our analysis also indicates that long and short wavelengths are stable.

\begin{figure}[t]
\centering
\includegraphics[width=1.0\linewidth]{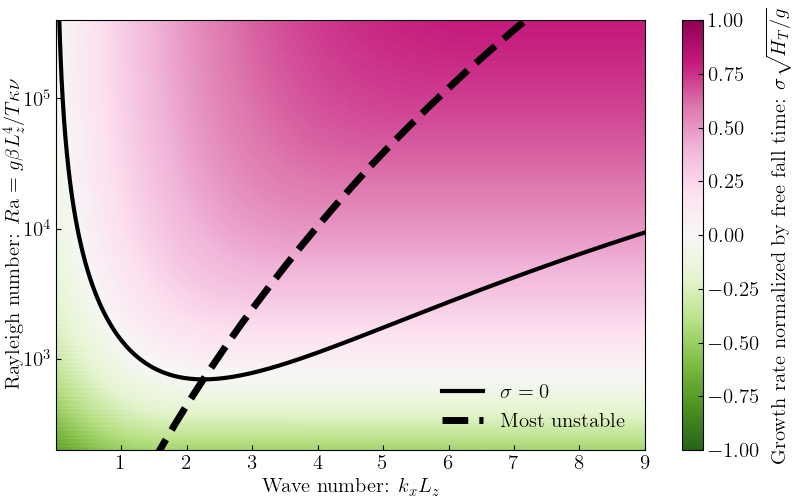}
\caption{The relation between $k_x$ and Rayleigh number $R{\rm a}$ with the growth rate derived by WKB approximation in the case of $H_T / L_z = 2$ and $\nu / \kappa = 1$.
The solid line shows the same line of the red dashed line in Figure \ref{fig:kRa_relation}, while the dashed line indicates the most unstable wave number for a given Rayleigh number $R{\rm a}$.}
\label{fig:kRa_relation_wSigma}
\end{figure}

Now, we can reflect growth rate derived by WKB approximation on this plane.
Figure \ref{fig:kRa_relation_wSigma} indicates $k_x - R{\rm a}$ relation with growth rate of each Rayleigh number in the case of $H_T / L_z = 2$ and $\nu/\kappa = 1$.
Note that the growth rate is normalized by free fall time over temperature scale height$\sqrt{H_T / g}$, which is different from other figures.
The solid line shows the same line of the red dashed line in Figure \ref{fig:kRa_relation}.
We find that the most unstable wavelength, represented by the dashed line in Figure \ref{fig:kRa_relation_wSigma}, differs from the wavelength associated with the critical Rayleigh number identified by Chandrasekhar \cite{Chandrasekhar1981}. 
The latter corresponds to the wavelength at the minimum of the black dashed line.

\subsection{Re-evaluation of critical Rayleigh number}
\label{subsec:HT_Racrit_relation}
In this section, we focus on the differences between the numerical solution and the analytical solution shown in Section \ref{sec:results}. 
In particular, in the case of $H_T / L_z = 2$, the WKB solutions consistently overestimate the growth rate compared to the numerical solution, while in the case of $H_T / L_z = 100$, they are in agreement with the numerical solutions.
The second-order derivative with respect to $z$ introduces viscosity artificially. 
However, a similar tendency is observed even in the absence of thermal conduction and viscosity or when only thermal conduction is considered. 
This suggests that such behavior is an intrinsic characteristic of the system and can generally be anticipated.
Therefore, it seems reasonable to suggest that similar differences may exist when viscosity is considered.

A comparison of the cases $H_T / L_z = 2$ and $H_T / L_z = 100$ illustrates the importance of reexamining the critical Rayleigh number \cite{Chandrasekhar1981, Turcotte2014} as a function of the temperature scale height.
The critical Rayleigh number is identified as the minimum value of the curve in Figure \ref{fig:kRa_relation}. 
If the Rayleigh number is below this value, convection is stabilized at all wavelengths, as shown in Figure \ref{fig:kRa_relation_wSigma}.
However, this criterion is derived through the WKB approximation, and it has been found that WKB solutions tend to overestimate the growth rate compared to numerical solutions, especially when the temperature scale height is of a similar magnitude to the domain height.
Thus, it is necessary to examine the critical Rayleigh number as a function of the temperature scale height using the numerical method.

\begin{figure}[ht]
\centering
\includegraphics[width=\linewidth]{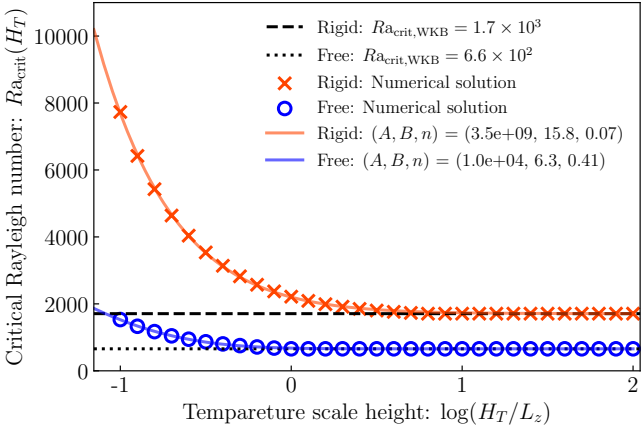}
\caption{
Critical Rayleigh number as a function of temperature scale height, $H_T$. 
Markers represent the critical Rayleigh number for each temperature scale height, 
while the solid lines indicate fitting curves based on Equation (\ref{eq:critRa_fitting}): $R{\rm a}_{\rm crit}(H_T) = R{\rm a}_{\rm crit, WKB} + A \exp [-B(H_T / L_z)^n]$. 
The colors distinguish the boundary conditions: red corresponds to rigid boundaries, and blue corresponds to free surface boundaries. 
The black dashed and dotted lines denote the values derived from the WKB approximation for rigid and free surface boundaries, respectively\cite{Chandrasekhar1981}.
}
\label{fig:discuss_HT_Racrit}
\end{figure}

Figure \ref{fig:discuss_HT_Racrit} illustrates the relationship between the temperature scale height, $H_T$, and the critical Rayleigh number, $R{\rm a}_{\rm crit}$. 
The results are fitted using the following trial function:
\begin{align}
    R{\rm a}_{\rm crit} ( H_T) = 
    R{\rm a}_{\rm crit, WKB} + A \exp \left [ -B \left ( \dfrac{H_T}{L_z} \right )^n \right ],
    \label{eq:critRa_fitting}
\end{align}
where $R{\rm a}_{\rm crit, WKB}$ is the value derived from the WKB approximation \cite{Chandrasekhar1981}, $A$ and $B$ are dimensionless fitting parameters, and $n$ is the power of $H_T/L_z$. 
For rigid boundaries, the parameters are $(A,B,n) \approx (3.5 \times 10^9,\ 15.8,\ 0.07)$, while for free surface boundaries, they are $(A,B,n) \approx (1.0 \times 10^4,\ 6.3,\ 0.41)$.

We find that the critical Rayleigh number is significantly larger than the estimate provided by the WKB approximation, especially when the temperature scale height is comparable to or smaller than the domain height, i.e., $R{\rm a}_{\rm crit}(H_T \lesssim L_z) \gg R{\rm a}_{\rm crit, WKB}$.
On the other hand, when the temperature scale height is much larger than the domain height, the numerical solutions converge to the WKB solutions.

\subsection{Verification of Boussinesq approximation}
\label{subsec:comparison_Boussinessq}
In this section, we apply our analysis to the Boussinesq approximation by modifying two key aspects: we replace the compressible continuity equation with its incompressible counterpart, and we replace the buoyancy term $g \delta \rho$ with $\rho_0 \alpha g \delta T$.
Where $\alpha$ is the coefficient of thermal expansion, defined as:
\begin{align}
\alpha = -\dfrac{1}{\rho} \left( \dfrac{\partial \rho}{\partial T} \right)_P,
\label{eq:Thermal_Coefficient}
\end{align}
which simplifies to $1/T$ for an ideal gas.
With these modifications, the following ordinary differential equations (ODEs) are derived for the continuity equation and the equation of motion in the $z$-direction:
\begin{align}
&
\dfrac{d\delta v_z}{dz} = -ik_x \delta v_x,
\label{eq:boussinesq_EoC}
\\[1mm]
&
\dfrac{d\delta P}{dz} =
-(\sigma + \nu k_x^2 ) \rho_0 \delta v_z - \rho_0 \alpha g\delta T.
\label{eq:boussinessq_EoMz}
\end{align}
The other ODEs remain unchanged from the previous analysis. 
With these modifications, we can apply our method to study the Boussinesq approximation.
This allows us to directly compare the results and assess the impact of the Boussinesq approximation on the dispersion relation.

\begin{figure*}[ht]
    \centering
    \includegraphics[width=0.9\linewidth]{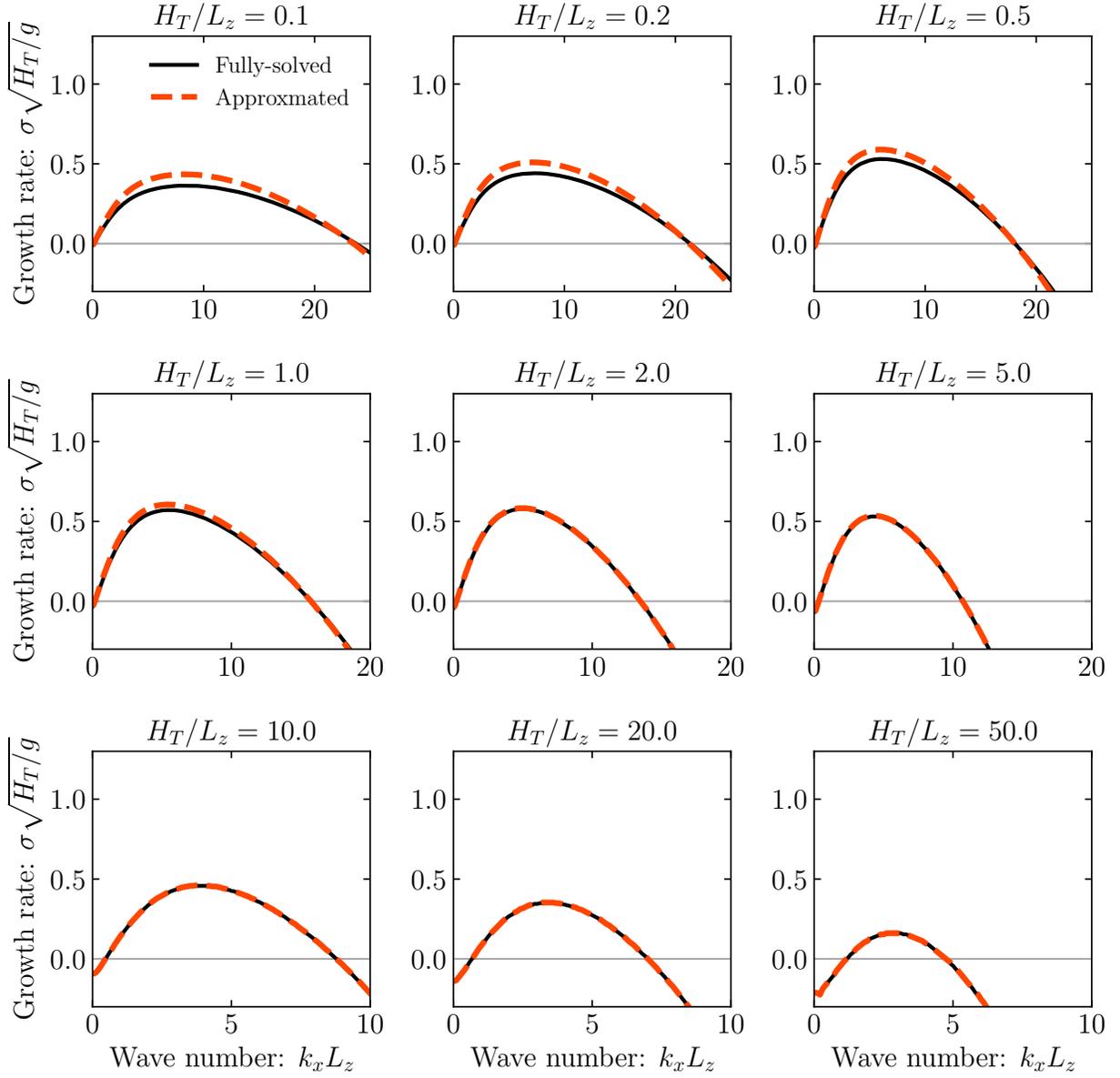}
    \caption{Dispersion relations calculated by varying the temperature scale height.
    The red dashed lines correspond to the Boussinesq approximation, while the black solid lines represent the non-approximated case.}
    \label{fig:Eigen_Comparison}
\end{figure*}

Figure \ref{fig:Eigen_Comparison} shows the comparison between cases with and without Boussinesq approximation.
Note that the growth rate is normalized by free fall time over temperature scale height, $\sqrt{H_T / g}$.
By varying the temperature scale height $H_T$ while keeping $\kappa / \sqrt{gL_z^3} = \nu / \sqrt{gL_z^3} = 10^{-2.5}$, we apply the shooting method using the same boundary conditions as described in Section \ref{sec:numerical}.

According to Figure \ref{fig:Eigen_Comparison}, when $H_T / L_z \gg 1$, the Boussinesq approximation matches the non-approximated case.
This finding suggests that the Boussinesq approximation is valid when the sound crossing time over the domain height, $t_{\rm snd} \equiv L_z / C_{\rm s}$, is much smaller than the free-fall time over the temperature scale height, $t_{\rm ff} \equiv \sqrt{H_T/g}$, i.e., $t_{\rm snd} / t_{\rm ff} \ll 1$.  
Rearranging this condition in terms of the temperature and pressure scale heights, $H_T$ and $H_P$, we obtain $(t_{\rm snd}/t_{\rm ff})^2 = (L_z/H_T)(L_z/\gamma H_P) \ll 1$ based on Equation \eqref{eq:intro:HP}. 
Moreover, according to Equation \eqref{eq:intro:HP_sum}, since $L_z / H_P > L_z / H_T$, the inequality  $(t_{\rm snd}/t_{\rm ff})^2 > \gamma^{-1} (L_z / H_T)^2$
is satisfied. 
Figure \ref{fig:Eigen_Comparison} indicates that the Boussinesq approximation is valid when the temperature scale height exceeds this threshold, i.e., $H_T / L_z \gg 1$.  
Note that this fact does not justify the WKB approximation, and we have not used the WKB approximation.

Conversely, in the case of $H_T / L_z \le 1$, we can see the discrepancies between the full solution and Boussinesq approximation, especially around the most unstable wavenumber.
These differences suggest that the Boussinesq approximation may not be fully accurate when the sound crossing time is comparable to or larger than the free-fall time, i.e., $t_{\rm snd}/t_{\rm ff} \ge 1$ or $H_T /L_z \lesssim 1$.
On the other hand, the growth time of convection is long near the critical wavelengths.  
Therefore, $t_{\rm snd} \ll t_{\rm growth}$, and the Boussinesq approximation is always valid.

We should remember that the Boussinesq approximation overestimate the growth rates in the case of $H_T / L_z \le 1$.
This discrepancy arises because the effect of pressure perturbation is neglected in the buoyancy term ($\delta \rho / \rho_0 = -\delta T / T_0$) in the Boussinesq approximation.
In reality $\delta \rho / \rho_0 = \delta P / P_0 -\delta T / T_0$, and $\delta P$ and $\delta T$ have different signs in the convective motion, and thus it overestimates the buoyancy force and the growth rate. 
However, the differences were not so pronounced within the parameter range considered in Fig. \ref{fig:Eigen_Comparison}.

\section{Summary}
\label{sec:summary}
In this section, we summarize the key findings of this study. 
This research investigates the situation in which the temperature and pressure scale heights are smaller than the domain height, with the aim of understanding the stabilization mechanisms of convection and reevaluating critical parameters, such as the Rayleigh number, in relation to the temperature scale height.  

The key findings are summarized as follows:
\begin{enumerate}
    \item \textbf{WKB Approximation:} 
    The WKB approximation accurately predicts the growth rate of convection when the temperature scale height is significantly larger than the domain height. 
    However, it overestimates the growth rate when the temperature scale height is comparable to the domain height, leading to overestimations of unstable wavelength ranges and the underestimation of the critical Rayleigh number.
    We do not find any overstabe mode of convection even in the case of small scale heights.
    
    \item \textbf{Boussinesq Approximation:}
    The Boussinesq approximation is effective when the sound crossing time over the temperature scale height is much smaller than the free-fall time over the temperature scale height.
    In contrast, a moderate amount of discrepancies arise when the sound crossing time is comparable to the free-fall time, but the difference is small within the parameter range considered in Fig. \ref{fig:Eigen_Comparison}.
    
    \item \textbf{Role of Two Kinds of Diffusion Terms:} 
    Only one diffusion cannot completely stabilize the system of negative entropy gradient.
    Thermal conduction only reduces the entropy perturbations, while viscosity only slows down the motion of fluid elements. 
    The combination of both diffusion mechanisms is required for complete stabilization of convection in negative entropy gradient structure that is driven by the coupling of the vortex perturbation, $\delta \omega_y$, and the temperature perturbation, $\delta T$.
    
    \item \textbf{Critical Rayleigh Number:}
    A refined estimation of the critical Rayleigh number, numerically derived using the shooting method, demonstrates that the WKB approximation tends to underestimate this value, when the temperature scale height is less than the domain height. 
    We provide a new approximate formula for the critical Rayleigh number.
\end{enumerate}

We hope this study provide an insight into the stability of the case of small temperature scale height and a step toward further investigations.

\begin{acknowledgments}
\end{acknowledgments}
We thank Hiroshi Kobayashi for useful discussion. 
This work was supported by JSPS KAKENHI Grant Number 18H05436, 18H05437 and 24KJ1302.

\nocite{*}
\bibliography{aipsamp}

\appendix
\section{Mode Coupling by Gravity}
\label{sec:couple}
In this section, we discuss the mode coupling by gravity, as outlined in Section \ref{subsec:contribution_diffusion}.

Considering a uniform medium without gravity and a temperature gradient, Equations (\ref{eq:discuss_vortex}) and (\ref{eq:discuss_temp}) can be transformed as follows:
\begin{align}
    &
    \sigma \delta \Omega_y = - \nu k^2 \delta \Omega_y,
    \label{eq:append_vortex}\\[1mm]
    &
    \sigma \delta T = - \kappa k^2 \delta T.
    \label{eq:append_temp}
\end{align}
Then, we can derive the following dispersion relation from these equations:
\begin{align}
    (\sigma + \nu k^2)(\sigma + \kappa k^2) = 0.
    \label{eq:append_DR}
\end{align}
The fact that Equations (\ref{eq:append_vortex}) and (\ref{eq:append_temp}) are independent of each other and that the dispersion relation can be factorized as shown in Equation (\ref{eq:append_DR}) implies that the diffusion of vortex and the diffusion of thermal fluctuation are independent of each other.
Equation (\ref{eq:append_vortex}) indicates the damping of the vortex mode (transverse wave): $\sigma = -\nu k^2$, while Equation (\ref{eq:append_temp}) represents the damping of the thermal fluctuation (longitudinal wave): $\sigma = -\kappa k^2$.

On the other hand, when we consider gravity, these modes are coupled as described by Equations (\ref{eq:discuss_vortex}) and (\ref{eq:discuss_temp}).
We can also confirm this by deriving the dispersion relation, which cannot be factorized, as follows:
\begin{align}
    (\sigma + \nu k^2)(\sigma + \kappa k^2) = \dfrac{g}{H_T} \dfrac{k_x^2}{k^2}.
    \label{eq:append_DR_wGrav}
\end{align}
Thus, this explains how the vortex mode and the thermal fluctuation are coupled by gravity.

\end{document}